%% file: main.tex
\documentclass[fleqn,usenatbib]{mnras}

\usepackage{newtxtext,newtxmath}

\DeclareRobustCommand{\VAN}[3]{#2}
\let\VANthebibliography\thebibliography
\def\thebibliography{\DeclareRobustCommand{\VAN}[3]{##3}\VANthebibliography}

\usepackage[utf8]{inputenc}
\usepackage[T1]{fontenc}
\usepackage{lipsum,hyperref,subfig}
\hypersetup{ 
			colorlinks=true,
			pdfauthor={Lamman, Claire},
			pdftitle={desiIA-draft},
			citecolor=blue
			}

\usepackage{graphicx}	
\usepackage{amsmath}	
\usepackage{lineno}
\let\oldequation\equation
\let\oldendequation\endequation
\renewenvironment{equation}
  {\linenomathNonumbers\oldequation}
  {\oldendequation\endlinenomath}
\usepackage[dvipsnames]{xcolor}
\usepackage{orcidlink}

\newcommand{\degree}{\circ} 

\title[Intrinsic Alignment as an RSD Contaminant]{Intrinsic Alignment as an RSD Contaminant in the DESI Survey}

\author[Claire Lamman]{
Claire Lamman \orcidlink{0000-0002-6731-9329},$^{1}$\thanks{E-mail: claire.lamman@cfa.harvard.edu}
Daniel Eisenstein,$^{1}$
Jessica Nicole Aguilar,$^{2}$ 
David Brooks,$^{3}$
Axel de la Macorra,$^{4}$\newauthor
Peter Doel,$^{3}$
Andreu Font-Ribera \orcidlink{0000-0002-3033-7312},$^{5}$
Satya Gontcho A Gontcho,$^{2}$ 
Klaus Honscheid,$^{6,7}$ 
Robert Kehoe,$^{8}$\newauthor
Theodore Kisner \orcidlink{0000-0003-3510-7134},$^{2}$ 
Anthony Kremin \orcidlink{0000-0001-6356-7424},$^{2}$ 
Martin Landriau \orcidlink{0000-0003-1838-8528},$^{2}$ 
Michael Levi \orcidlink{0000-0003-1887-1018},$^{2}$ 
Ramon Miquel,$^{5,9}$\newauthor
John Moustakas \orcidlink{0000-0002-2733-4559},$^{10}$
Nathalie Palanque-Delabrouille \orcidlink{0000-0003-3188-784X},$^{2,11}$ 
Claire Poppett,$^{2,12,13}$ 
Michael Schubnell,$^{14,15}$ \newauthor
Gregory Tarlé \orcidlink{0000-0003-1704-0781}$^{15}$
\\ \\
$^{1}$Center for Astrophysics $|$ Harvard \& Smithsonian, 60 Garden Street, Cambridge, MA 02138, USA\\
$^{2}$Lawrence Berkeley National Laboratory, 1 Cyclotron Road, Berkeley, CA 94720, USA\\
$^{3}$Department of Physics \& Astronomy, University College London, Gower Street, London, WC1E 6BT, UK\\
$^{4}$Instituto de F\'{\i}sica, Universidad Nacional Aut\'{o}noma de M\'{e}xico,  Cd. de M\'{e}xico  C.P. 04510,  M\'{e}xico\\
$^{5}$Institut de F\'{i}sica de'Altes Energies (IFAE), The Barcelona Institute of Science and Technology, Campus UAB, 08193 Bellaterra Barcelona, Spain\\
$^{6}$Center for Cosmology and AstroParticle Physics, The Ohio State University, 191 West Woodruff Avenue, Columbus, OH 43210, USA\\
$^{7}$Department of Physics, The Ohio State University, 191 West Woodruff Avenue, Columbus, OH 43210, USA\\
$^{8}$Department of Physics, Southern Methodist University, 3215 Daniel Avenue, Dallas, TX 75275, USA\\
$^{9}$Instituci\'{o} Catalana de Recerca i Estudis Avan\c{c}ats, Passeig de Llu\'{\i}s Companys, 23, 08010 Barcelona, Spain\\
$^{10}$Department of Physics and Astronomy, Siena College, 515 Loudon Road, Loudonville, NY 12211, USA\\
$^{11}$IRFU, CEA, Universit\'{e} Paris-Saclay, F-91191 Gif-sur-Yvette, France\\
$^{12}$Space Sciences Laboratory, University of California, Berkeley, 7 Gauss Way, Berkeley, CA  94720, USA\\
$^{13}$University of California, Berkeley, 110 Sproul Hall \#5800 Berkeley, CA 94720, USA\\
$^{14}$Department of Physics, University of Michigan, Ann Arbor, MI 48109, USA\\
$^{15}$University of Michigan, Ann Arbor, MI 48109, USA\\
}

\pubyear{2022}

\begin{document}
\label{firstpage}
\pagerange{\pageref{firstpage}--\pageref{lastpage}}
\maketitle

\begin{abstract}
We measure the tidal alignment of the major axes of Luminous Red Galaxies (LRGs) from the Legacy Imaging Survey and use it to infer the artificial redshift-space distortion signature that will arise from an orientation-dependent, surface-brightness selection in the Dark Energy Spectroscopic Instrument (DESI) survey. Using photometric redshifts to down-weight the shape-density correlations due to weak lensing, we measure the intrinsic tidal alignment of LRGs. Separately, we estimate the net polarization of LRG orientations from DESI's fiber-magnitude target selection to be of order $10^{-2}$ along the line of sight. Using these measurements and a linear tidal model, we forecast a 0.5\% fractional decrease on the quadrupole of the 2-point correlation function for projected separations of 40-80 $h^{-1} {\rm Mpc}$. We also use a halo catalog from the A\textsc{bacus} S\textsc{ummit} cosmological simulation suite to reproduce this false quadrupole.
\end{abstract}




\begin{figure*}
\includegraphics[scale=.25]{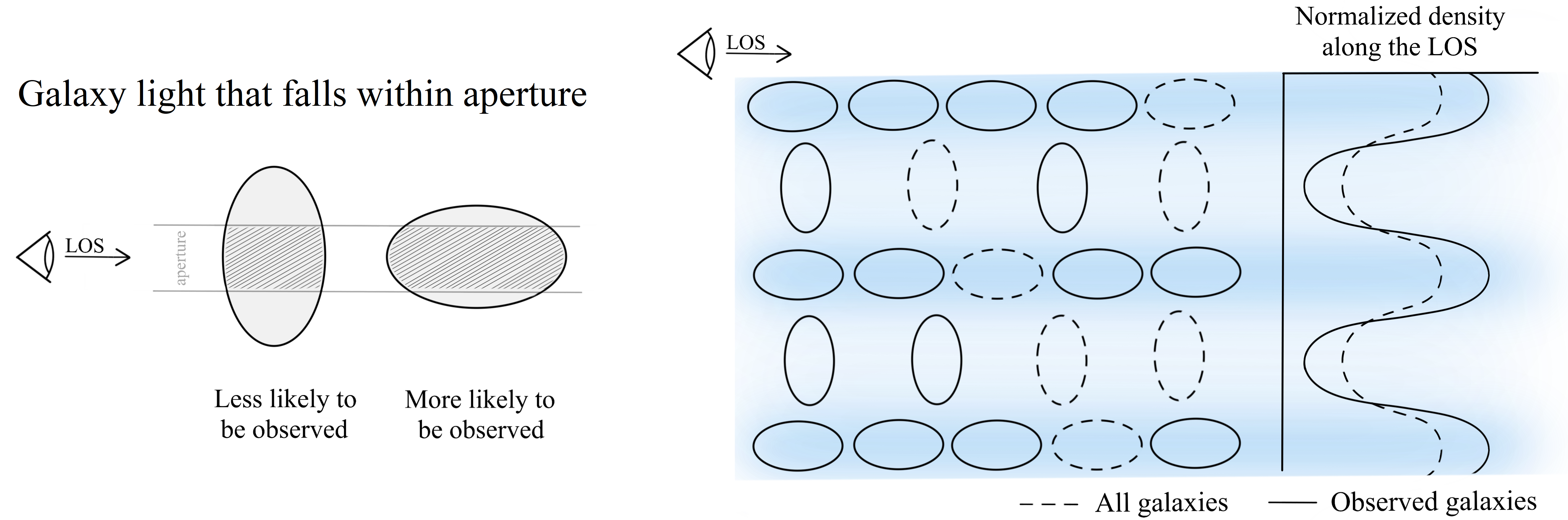}
\caption{A cartoon demonstrating how an aperture-based selection can combine with tidal alignment to affect measurements of the underlying density. Elliptical galaxies will have the maximum concentration of light on the sky when their primary axis is pointed at the observer. In this case, more of the light falls within an aperture and it is more likely to be included in DESI's fiber magnitude selection. The cartoon on the right shows galaxies with maximum tidal alignment lying along density filaments which are parallel to the LOS. These filaments are represented with a blue gradient. Galaxies in filaments tend to be oriented in the direction of the filament, and ones between tend to point toward the higher density regions. In this case, DESI is more likely to select galaxies in denser regions, resulting in an amplification of this density mode. The opposite effect happens for filaments which are perpendicular to the LOS (not shown here, see Figure 1 of \citet{martens_radial_2018}). Since DESI is more likely to select galaxies in filaments which lie along the LOS, and less likely to select ones in perpendicular filaments, an anisotropic clustering arises and biases the RSD signal.}
\label{fig:bias_diagram}
\end{figure*}

\section{Introduction}
Redshift-Space Distortions (RSD), are an effect often used for measuring the growth of large-scale structure. On the scale of galaxy clusters, peculiar velocities of galaxies ``smear" structure along the line of sight (LOS) in redshift space \citep{jackson_critique_1972}. On larger scales, material falling into over-dense regions creates a ``squashing" effect along the LOS \citep{kaiser_clustering_1987}. The difference in clustering along versus transverse to the LOS can be described by the quadrupole of the correlation function, $\xi_2$. This needs to be corrected for to map galaxies in real space, and on large scales is a measurement of the growth rate of structure and can be used to test gravity. \par

To fully utilize RSD measurements in large spectroscopic galaxy surveys, one of their important biases must be understood: intrinsic galaxy alignment (IA). The primary axis of galaxies can be intrinsically aligned with each other (II correlation) and with the underlying density, or tidal field (GI correlation). When a galaxy survey has an orientation-dependent selection bias and galaxy orientations are also correlated with the tidal field, $\xi_2$ is directly affected.\par

\cite{hirata_tidal_2009} used linear models of tidal alignment and orientation-dependent selections to predict that GI correlations could affect RSD measurements by as much as 10\%. This effect is highly survey-dependent due to its strong dependence on survey selection and the differences in tidal alignments between galaxy samples. \cite{martens_radial_2018} and \cite{obuljen_detection_2020-1} have measured an anisotropic galaxy assembly bias in the Baryon Oscillation Spectroscopic Survey (BOSS). Since the velocity dispersion of elliptical galaxies is non-isotropic and may correlate with axis orientation and tidal environment, this effect could be a manifestation of the effect described by \cite{hirata_tidal_2009}. On the other hand, \cite{singh_fundamental_2021} followed a similar method to \cite{martens_radial_2018} and found the Fundamental Plane of BOSS galaxies to be dominated by systematics and poorly correlated with IA, resulting in a null detection of the RSD IA bias for BOSS. \par

As a Stage IV survey, it is necessary to not only detect, but quantify these biases for the Dark Energy Spectroscopic Instrument (DESI). DESI is in the midst of a 5-year survey, measuring spectra of over 40 million galaxies within 16,000 deg$^2$ of the sky \citep{desi_collaboration_desi_2016, abareshi_overview_2022}. 

Successful inference of a galaxy's spectroscopic redshift depends on target surface brightness. This is especially true for a large survey like DESI, which prioritizes survey speed at the cost of higher signal-to-noise. To impose this explicitly, DESI adopts a surface brightness-dependent cut: limiting the magnitude within an aperture instead of the objects' total magnitude. While this mitigates systematic errors related to surface brightness, it creates a bias in the 3D orientation of galaxies. Galaxies with a pole-on orientation have a higher surface brightness and are more likely to be selected. Since galaxies with tidal alignments tend to point towards regions of higher density, this can also mean preferentially selecting galaxies which lie in filaments along the LOS. This results in an enhancement of clustering in the radial direction and suppression in the transverse direction, mimicking RSD. The key piece of modeling this effect is relating the polarizability of the surface brightness selection to the shape of galaxies viewed from "the side", i.e. transverse to the LOS. This depends on the details of the light profiles and triaxial shapes of the galaxies (Figure \ref{fig:bias_diagram}). 

About 20\% of DESI's targets are Luminous Red Galaxies (LRGs), which fall in the redshift range $0.4-1.0$ \citep{zhou_clustering_2021}. These high-mass, relatively inactive galaxies exhibit large tidal alignments \citep{hirata_intrinsic_2007} and are more affected by an aperture-based selection because they have larger angular sizes than Emission Line Galaxies (ELGs). Therefore, we chose to focus our investigation on LRGs as the DESI sample most likely to be substantially biased by these alignments, although our methods would also work for ELGs. \par

The two effects that combine to create this bias, GI alignment and selection-induced polarization, can both be estimated and used to calibrate the quadrupole $\xi_2$. Here we measure the shape-density correlation of LRGs as projected on the plane of the sky using shapes from the DESI Legacy Imaging Survey \citep{dey_overview_2019}. We isolate the signal of intrinsic positions from weak lensing via photometric redshifts, model DESI's orientation-dependent selection function, and put our detection in context of $\xi_2$ via a linear tidal model. As an additional test, we use the A\textsc{bacus}S\textsc{ummit} cosmological simulations to reproduce an aperture-based selection and measure the effect on $\xi_2$.

\section{DESI Catalogs}\label{sec:desi_catalog}

\subsection{Imaging}\label{sec:imaging}
Our measurements of GI alignment were made with LRGs from the Legacy Imaging Survey, DR9 \citep{dey_overview_2019}. This is the catalog DESI uses to select its targets, and contains imaging in three bands (g, r, and z) and projected shapes for sources in $14,000\deg^2$ of the extra galactic sky. It also includes photometry from the Wide-field Infrared Survey Explorer, which contains $r$ and $W1$ fluxes that are corrected for Milky Way extinction. The LRG target selection includes a cut based on the expected flux which falls within a DESI fiber. The $z$-band magnitude within a 1\arcsec.5 - diameter aperture is limited to $z_{\text{fiber}}<21.61$ in the Northern Galactic Cap and $z_{\text{fiber}}<21.60$ in the Southern Galactic Cap. For more information on the photometric selection of DESI's LRG sample, see \citep{zhou_target_2022}.\par

The source of each target (after deconvolving with a point spread function) is modeled as several light profiles at the pixel level using T\textsc{ractor} \citep{lang_tractor_2016}. Based on the fits' $\chi^2$ values, we used shape parameters from the best fit out of these models: exponential disk, de Vaucouleurs, and Sersic. This is different from DESI's default selection, which includes PSF and round-exponential fits, and a marginalized $\chi^2$ criteria to avoid over-fitting bright targets as round-exponentials. These models were avoided for our measurements, as circles have no distinguishable orientation. \par 

Quality cuts were applied to target declinations $\delta >-30 ^{\circ}$ and galactic latitudes $b >20 ^{\circ}$.  $r-W1$ color correlates well with redshift, so we used this color for the pair selection and weighting scheme detailed in \ref{sec:weighting}. To conform with these weights, color outliers were removed by requiring $1 < r-W1 < 4.5$. Our final sample contained 17.5 million LRGs.\par

\subsection{Spectroscopy}\label{desi_redshifts}
We calibrate our photometric redshifts using a large sample of spectroscopic redshifts from the DESI Survey Validation (SV) observations. SV is designed to represent the full survey and is used to assess DESI's target selection. We use DESI's internal SV catalog, Fuji, which comprises of quality observations taken from 14 December 2020 through 10 June 2021. From this we selected 133,924 LRGs with colors $0.6<r-z$ and $1.5<r-W1<4.5$, and redshifts $0.001<\text{z}<1.4$.

\section{Intrinsic Alignment Signal}\label{sec:alignment}

\subsection{Alignment Formalism}\label{formalism}

\begin{figure} 
\includegraphics[scale=0.5]{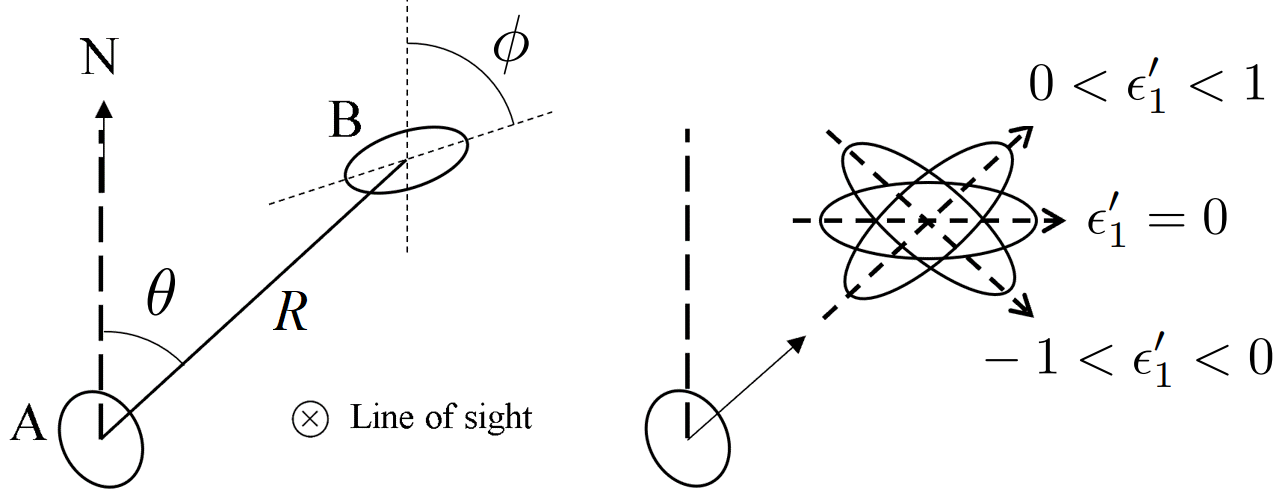}
\caption{How angles are defined in our alignment metric. For a given pair of galaxies,  $\epsilon_1'$ is maximum when the axis of B is parallel to the separation vector between it and A (most aligned), and minimum when it's perpendicular (anti-aligned). $\epsilon'_1$ also scales with axis ratio; it approaches 0 as the shape of B becomes more circular. This is measured as a function of transverse separation, $R$.}
\label{fig:formalism_diagram}
\end{figure}

The projected alignment of galaxies on the sky is quantified with a relative complex ellipticity (Figure \ref{fig:formalism_diagram}). This measures the degree to which a galaxy is aligned with, and stretched along, a separation vector between it and another galaxy. Measuring this as a function of the separation vector's magnitude, $R$, for many galaxy pairs is a way to quantify the alignment of LRGs to the underlying tidal field.\par

Here, 2D galaxy shapes are modeled as ellipses with a complex ellipticity
\begin{equation}
    \epsilon = \frac{a-b}{a+b}\exp{2i\phi}
\end{equation}
where $a$ and $b$ are the primary and secondary axis of the 2D ellipse, and $\phi$ is the orientation angle of the primary axis, measured East of North. We define the ellipticity of a galaxy $B$ relative to another galaxy $A$ using the difference between $B$'s orientation angle, $\phi_B$, and its position angle relative to $A$, $\theta_{BA}$, also measured East of North.
\begin{equation}
    \theta_{BA}' = \phi_B - \theta_{BA}
\end{equation}
This gives us a relative ellipticity, for which we measure the real component:
\begin{equation}
    \epsilon_{BA}' = \frac{a_B-b_B}{a_B+b_B}\exp{2i\theta_{BA}'} \\
\end{equation}
\begin{equation}\label{eq:e1}
    \epsilon'_1 = \text{Re}(\epsilon') =  |\epsilon'|\cos{2\theta'} \\
\end{equation}
This measurement is averaged over many pairs of galaxies as a function of their angular separations on the sky to obtain $\mathcal{E}(R)$, the 2D shape-density correlation.

\begin{figure} 
\includegraphics[scale=0.37]{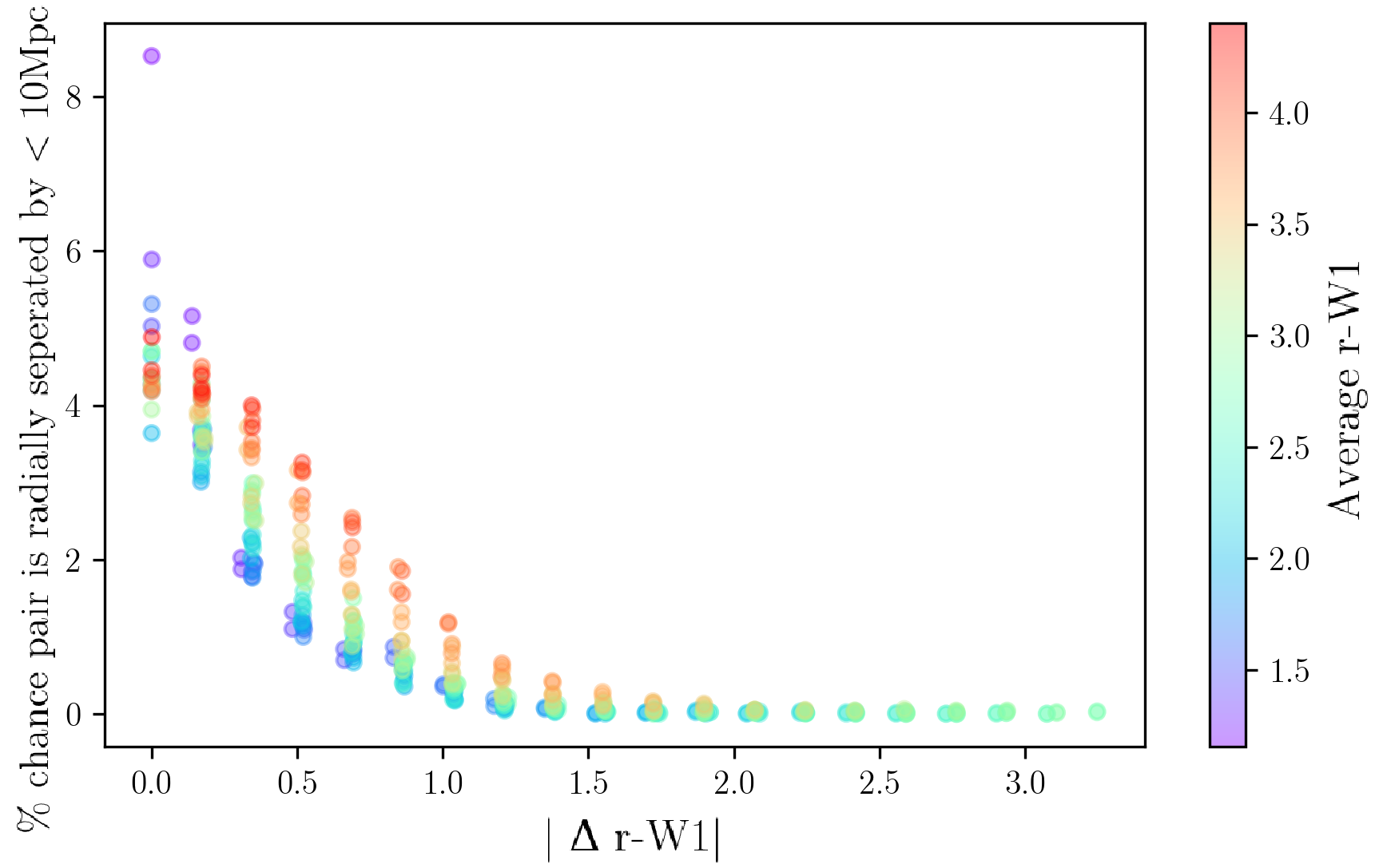}
\caption{Visualization of the matrix used for weighting pairs of galaxies in the alignment signal based on their color difference. This weighting scheme was created using a sub-sample of DESI galaxies with measured redshifts and favors pairs which are more likely to be physically associated with each other.}
\label{fig:lookup_matrix}
\end{figure}

\subsection{Color Weighting}\label{sec:weighting}
As our signal is a function of transverse separation, the main source of its dilution is from pairs of galaxies with large separations along the line of sight. At the time of this paper, we do not have spectra for all of the imaged galaxies and so use color as a redshift proxy. To give pairs which are more likely to be physically associated a higher weight in the alignment signal, we created a weighting scheme based off of their $r-W1$ colors. \par
This scheme gives higher weights to galaxies which are more likely to have small separations along the line of sight. For a pair of galaxies with two colors, we used existing redshifts to estimate the likelihood that they were separated by less than $10 {\rm Mpc}$. Using the redshifts DESI has measured so far, described in Section \ref{desi_redshifts}, we separated galaxies into 20 bins of $r-W1$ color. For every combination of the average colors in each bin, we estimated the fraction of galaxies which are radially separated by less than $10\ {\rm Mpc}$, based on their redshift difference and assuming the Hubble flow. The resulting lookup matrix was then used as a weight when averaging the alignment signal from individual pairs (Figure \ref{fig:lookup_matrix}).

\subsection{Intrinsic Alignment Measurement}
The catalog was divided into 10 groups based declination and then each of those into 10 groups based on right ascension, resulting in 100 sky regions with an equal number of galaxies in each, ~1.8 million. We measured the projected alignment of neighboring galaxies relative to each galaxy in each region. This was averaged over 20 bins of transverse, angular separation $R$, resulting in 100 determinations of the IA signal. The average and standard error of these 100 measurements at each separation is our projected IA measurement, $\mathcal{E}(R)$\footnote{code available here: \href{https://github.com/cmlamman/ellipse_alignment}{github.com/cmlamman/ellipse\_alignment}}.

\begin{figure} 
\includegraphics[scale=0.23]{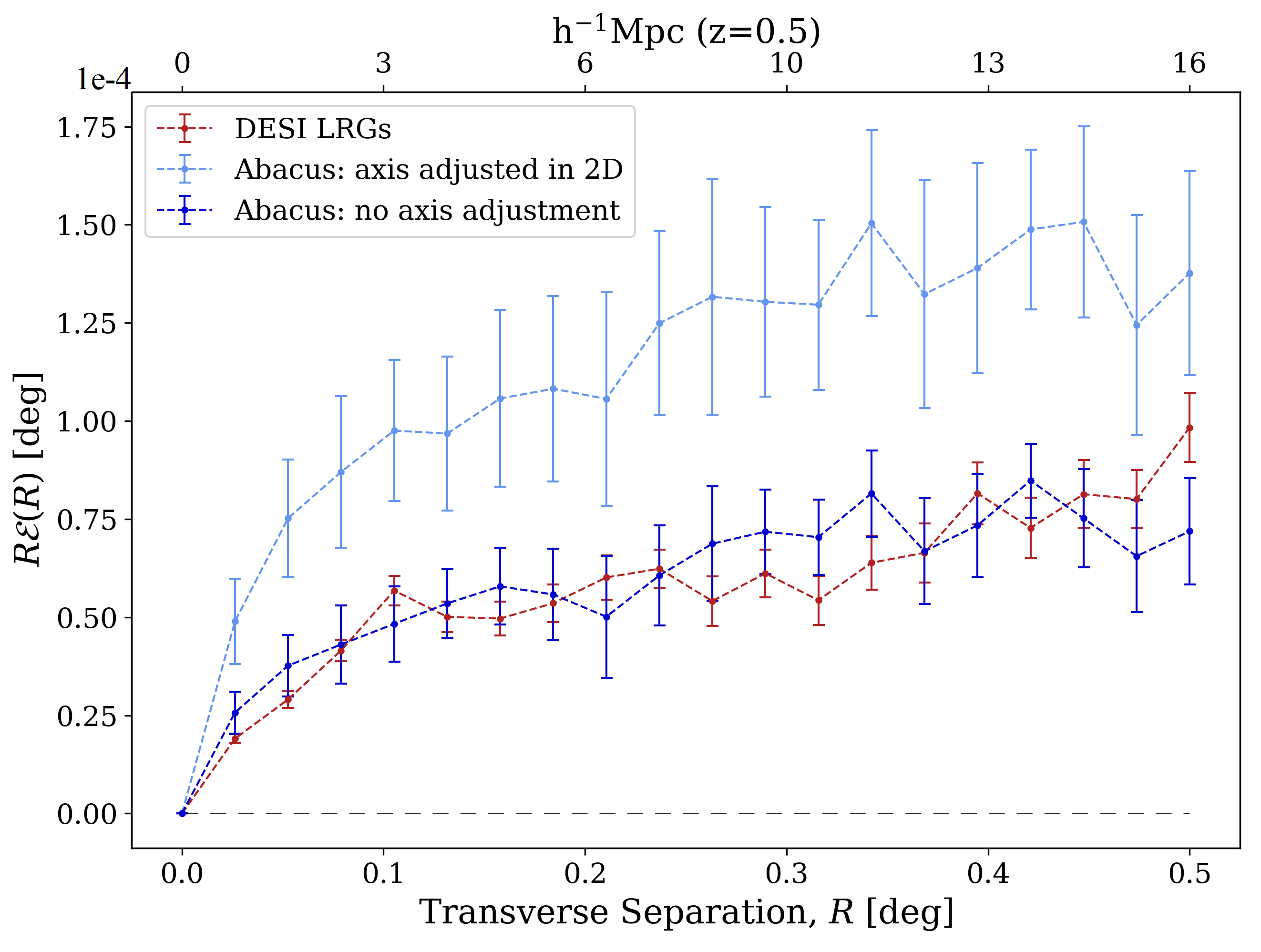}
\caption{Our final measurement of the projected shape-density correlation of DESI LRGs (red), which includes weighting based on the color difference in galaxy pairs. This is compared to the same measurement made with Abacus halos (dark blue). The light blue line shows the alignment of Abacus halos once the distribution of their projected shapes was adjusted to match the LRGs, but does not include galaxy-halo misalignments.}
\label{fig:ia_main}
\end{figure}

\begin{figure}
\includegraphics[scale=0.25]{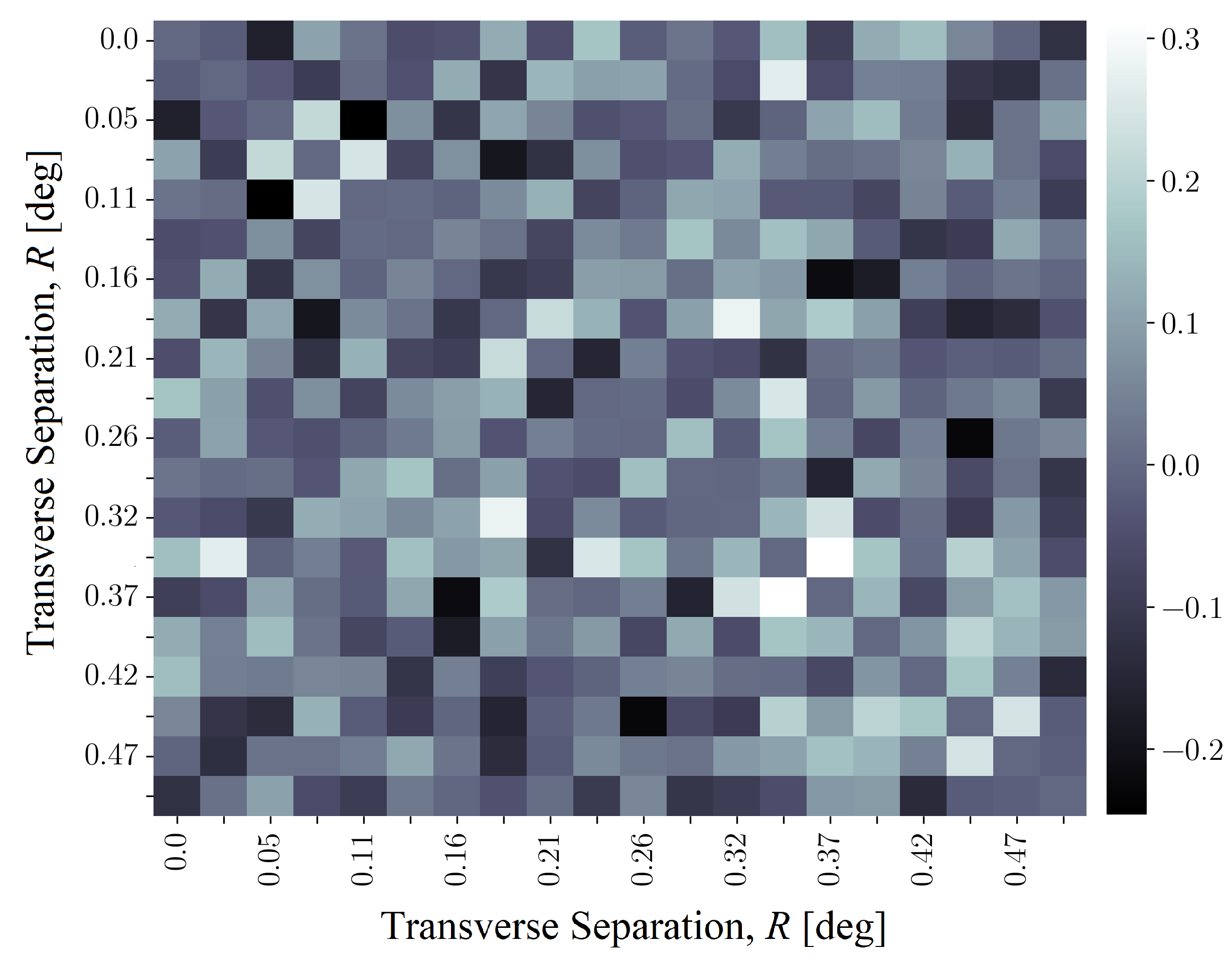}
\caption{The reduced covariance matrix of $\mathcal{E}$ between bins of transverse separation for our IA measurement; the identity matrix has been subtracted from this plot. This demonstrates that the measurements of $\mathcal{E}$ in each bin of transverse separation are statistically independent of each other.
}
\label{fig:ia_cov}
\end{figure}

Our final determination of $\mathcal{E}(R)$ for DESI LRGs is displayed in Figure \ref{fig:ia_main}. This signal broadly agrees with our measurement of projected IA in the Abacus Mock from Section \ref{sec:abacus}, which did not include any misalignments from the original halo orientations. The similarity between the alignment in LRGs and raw halo shapes is likely a coincidence due to two opposing effects: halo orientations are more aligned with the underlying density, which increases $\mathcal{E}$, but are rounder than LRGs, which dilutes $\mathcal{E}$. The LRG measurements of $\mathcal{E}$ in each angular bin are statistically independent of each other, as demonstrated by the covariance of our final $\mathcal{E}(R)$ signal between the 20 bins of transverse separation (Figure \ref{fig:ia_cov}). This indicates that there are no systematic errors in our shape measurements which are correlated with the underlying matter distribution.

\begin{figure*} 
\includegraphics[scale=.4]{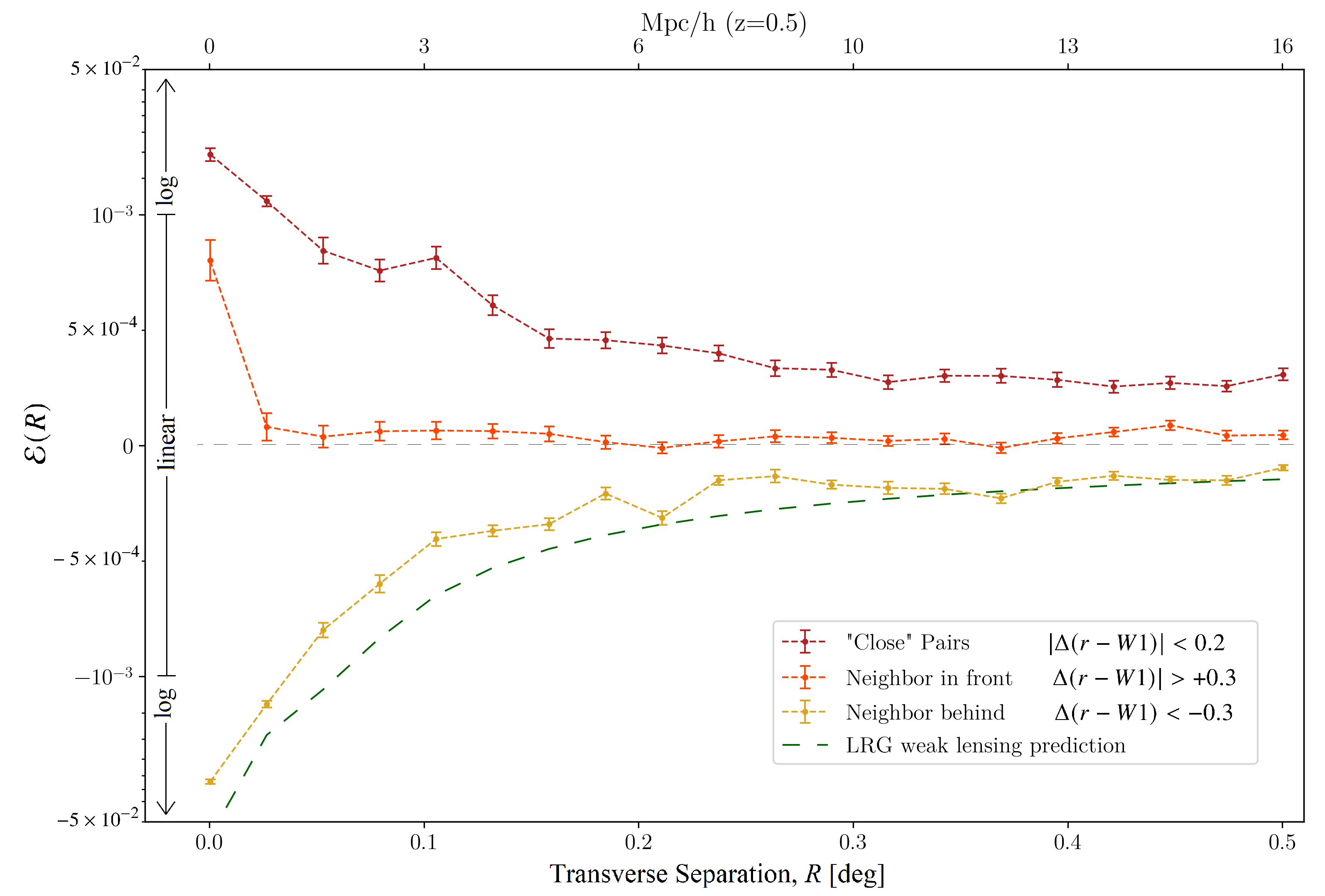}
\caption{The shape-density correlation of DESI LRGs, resulting from measuring the shape of one "neighbor" galaxy relative to the separation vector between it and another. The top abscissa displays the comoving distance corresponding to the transverse separation that was measured. No color weights were used for these measurements. Instead, measurements were made using different color-based restrictions for each pair of galaxies in order to explore the effects of weak lensing on the IA signal. The dark red line is the resulting signal when only measuring galaxy pairs which have a very similar $r-W1$ color, to approximate physical proximity. The orange and yellow lines are both measurements made on pairs of galaxies which have a large difference in $r-W1$ color, to emulate pairs which have no physical association. For the measurement shown in orange, we used pairs in which the neighbor galaxy was more blue than the other. Therefore we only measured the shape of galaxies relative to ones behind it, so their shapes were broadly unaffected by weak lensing. The converse was applied for the signal shown in yellow; here we only measured the shape of a galaxy if it was much redder than its counterpart. This means that the measured shape correlation is dominated by weak lensing.}
\label{fig:color_pairs}
\end{figure*}

\subsection{Weak Lensing}\label{lensing}
Besides intrinsic alignments, the main effect impacting the alignment signal is gravitational weak lensing. If the shape of a neighbor galaxy is measured relative to a foreground central galaxy, the neighbor's light can be gravitationally distorted by the central before reaching us. 

Since we only measure the shape of one galaxy in each pair, weak lensing is only present when the measured galaxy is behind the central one. Therefore a simple way to isolate the weak lensing and IA signals is to set restrictions on the radial separations of pairs. Using $r-W1$ color again as a distance proxy, we measured the alignment for sets of pairs with various color restrictions (Figure \ref{fig:color_pairs}). The signal from only measuring the shapes of galaxies relative to their closest color neighbors is comparable to our measurement using color weighting. The signal from only measuring galaxy shapes relative to background galaxies is consistent with $0$ above separations of a few Mpc, and the signal from measuring galaxy shapes relative to foreground galaxies is, as expected, opposite in sign to the intrinsic signal.\par

To check whether the lensing signal is consistent with expectations, we consider the following approximate model. The net effect of weak lensing acts in opposition to the IA signal, as it creates a tangential shear on the sky:
\begin{equation}
    \gamma_t = \frac{\bar{\Sigma}(<r_p) - \Sigma(r_p)}{\Sigma_{\rm crit}}
\end{equation}

\noindent where $\bar{\Sigma}(<r_p)$ is the average surface overdensity with some transverse distance  $r_p$. Assuming a power-law model for the correlation function $\xi_{gg} = (r_0/r)^2$, the surface overdensity at at projected separation $r_p$ is given as
\begin{equation}\label{eq:lens_2}
    \Sigma(r_p) = \pi\frac{\rho_0}{\beta}\frac{r_0^2}{r_p}
\end{equation}
And the average overdensity within $r_p$ is 
\begin{equation}\label{eq:lens_1}
    \bar{\Sigma}(<r_p) = \pi\frac{\rho_0}{\beta}\frac{r_0^2}{r_p}
\end{equation}
\noindent The derivations of these expressions for projected overdensity can be found in Appendix \ref{appendix:wl}. $\Sigma_{\rm crit}$ is the critical mean density, above which the light of a source is split into multiple images.
\begin{equation}
    \Sigma_{crit} = \frac{c^2 D_S}{4\pi G D_L D_{LS}}
\end{equation}

Here, $r_0 = 7.78\ {\rm Mpc}/ h$ is the 3D correlation length for DESI clustering \citep{kitanidis_imaging_2020}, $\beta = 2.15$ is the clustering bias for DESI LRGs \citep{zhou_clustering_2021}, and $\rho_0 = 2.68\times 10 ^{-30} \text{ g }\text{cm}^{-3}$ is the critical matter density of the Universe from Planck 2018 \citep{collaboration_planck_2020}. $D_S$, $D_L$, and $D_LS$ are the distances to the source, distance to the lens, and distance between them, respectively.\par

To connect this to our alignment formalism described in Section \ref{formalism}, the tangential shear is defined as 
\begin{equation}
    \gamma_t = \frac{a-b}{a+b} e^{2i\phi}
\end{equation}
\noindent where $\phi$ is the azimuthal angle of the source galaxy's primary axis with respect to the lens. This results in the relation
\begin{equation}
    \bar{\epsilon_1}' = \frac{\bar{\gamma}_t}{-2}
\end{equation}

We then estimated the amplitude of this signal in our sample. To obtain $D_S / D_L D_{LS}$, we used photometric redshifts for every pair of galaxies, and average the result. We used a simple, linear fit of our DESI spectroscopic sample to estimate these redshifts:
\begin{equation}
    \text{z} = 0.25(r\text{-}W1 \text{ color}) - 0.02
\end{equation}

The resulting lensing estimation is shown in Figure \ref{fig:color_pairs}. It agrees well with the IA measurement made when limiting to pairs we expect are only affected by lensing, though it is a simple estimate that did not go into our final results. The final IA signal is likely still diluted by weak lensing. However we did not develop a more sophisticated adjustment for lensing, as DESI's first year of spectra will allow us to sufficiently isolate physically-associated pairs.

 \section{IA with Abacus Mock Catalog}\label{sec:abacus}\label{mapping}\label{color_mock}
To contextualize the measured IA signal, we built a mock catalog from the A\textsc{bacus}S\textsc{ummit} C\textsc{ompa}SO halo catalog \citep{hadzhiyska_compaso_2021}. A\textsc{bacus}S\textsc{ummit} is a suite of large, high-accuracy, high-resolution cosmological simulations made with the A\textsc{bacus} N-body code \citep{maksimova_abacussummit_2021}. We used halos from a box with comoving $2000\ h^{-1}{\rm Mpc}$ sides, simulated at $z=0.725$.\par

We mapped the halos' comoving positions to redshift and sky coordinates by placing an observer $1700\ h^{-1}{\rm Mpc}$ away from the center of the box along the $x$-axis. To have an even sky distribution and consistent redshift range across the sky, we set boundaries of $\pm12^{\circ}$ in right ascension and declination, with a redshift range of $0.51<\text{z}<0.97$.\par
We then selected the largest halos to match both the LRG density of our DESI sample within this redshift range, $7.3\times10^{-4}\ (h^{-1}{\rm Mpc})^{-3}$, and the redshift distribution from DESI spectra. Our final mock catalog contains 766,341 halos.\par

To imitate the DESI Legacy Imaging Survey colors, we used a catalog of DESI LRG spectroscopic redshifts. They were sorted by redshift and each assigned an index. For each halo, we identified the LRG with the closest redshift percentile. We then smoothed our selection by sampling a neighboring LRG from a Gaussian of indices centered at the index of closest redshift and with a width of 300. After taking the $r-W1$ color from the LRG with the resulting index, we again smoothed by sampling a Gaussian centered at that color, with a width of $\sigma=0.03$. These smoothing parameters sufficiently reproduced the observed data spread, and variations of them do not significantly affect the measured alignment signal.\par

The A\textsc{bacus} 3D halo shapes are modeled as triaxial ellipsoids. A common method for finding the projected axis ratios of ellipsoids can be found in \citet{binney_testing_1985}. For measuring the alignment of galaxy shapes, we additionally need the orientation angle of the projected shape. Therefore, we adapted the method derived in \cite{gendzwill_analysis_1981} to project triaxial ellipsoids onto the celestial sphere. Our process for obtaining the axis ratio and orientation of an ellipsoid projected along an arbitrary viewing angle can be found in Appendix \ref{appendix:projection}.\par

Halos are rounder than LRGs, so we mapped the axis ratios of the projected halos to the LRG axis ratio distribution. We adjust each axis ratio, $b/a=d$, with the empirical function:
\begin{equation}
    d' = 1 + 1.1(d-1) - 2.035(d-1)^2+1.76(d-1)^3
\end{equation}
This function correctly reproduces the number of observed axis ratios which fall in 100 bins between 0 and 1. We made no adjustments for the orientations of halos.

Using the same color-weighting scheme as described in Section \ref{sec:alignment}, we measure the projected shape-density correlation of our resulting mock catalog. The result can be seen in Figure \ref{fig:ia_main}. The higher amplitude is likely due to the simulation not including the effects of weak lensing and the higher degree of alignment in halos compared to galaxies. \cite{tenneti_galaxy_2014} estimates large, central galaxies at DESI redshifts to be misaligned with their host halo by an average of around 10-20$^\degree$. Assuming random misalignment, this propagates to a $\mathcal{E}$ signal that is 75-94\% of the same signal without misalignments.

\section{Modeling alignment - \texorpdfstring{$\xi_2$}{} Correlation}\label{section:model}

\subsection{Linear Tidal Model}
We adopt a linear tidal model to connect IA and DESI's shape selection bias with the quadrupole of the correlation function, $\xi_2$. This approximation assumes that the projected shapes of galaxies are linearly related to the projected density distribution and holds for LRGs above projected separations of $10\ h^{-1} {\rm Mpc}$ \citep{catelan_correlations_2001, hirata_intrinsic_2004, singh_intrinsic_2015, troxel_intrinsic_2015}.\par
We define $\nabla^2 \phi = \rho$,
where $\rho$ is the fractional over density.
The tidal tensor is then the traceless combination 
$T_{ij} = \partial_i \partial_j \phi - (1/3) \delta^K_{ij} \nabla^2 \phi$,
where $\delta^K$ is the Kronecker delta.\par
We model the mean 3D ellipticity of a triaxial galaxy as $\tau T_{ij}$,
where the axis lengths behave as $I+\tau T$.
For this derivation, we assume that 2D projections of such galaxies behave as the 2D projection
of these lengths.  The mean eccentricity tensor must also be traceless, so
for a projection with $\alpha$, $\beta$ = $\{x,y\}$, the projected ellipticity is given as
$\epsilon_{\alpha\beta} = \tau(T_{\alpha\beta} + T_{zz}/2)$,
where we used $T_{xx}+T_{yy} = -T_{zz}$. \par
Using Fourier-space conventions, the tidal tensor model $T_{ij}$ can be expressed as
\begin{equation}
\begin{split}
T_{ij}(\vec r) = \left(\partial_i\partial_j -\frac{\delta^K_{ij}}{3}\nabla^2\right)
\int \frac{d^3k}{(2\pi)^3} \tilde\phi(\vec k)e^{i\vec k\cdot\vec r}\\
= \int \frac{d^3k}{(2\pi)^3} \left( \frac{k_i k_j - \delta^K_{ij}k^2/3}{k^2}\right) \tilde\rho(\vec k) e^{i\vec k\cdot\vec r}
\end{split}
\end{equation}

\subsection{Shape-Density Correlation}
To connect a bias in ellipticity and a projected shape-density correlation with a $\xi_2$ signature, we first consider how galaxy ellipticty correlates with surface density. We begin with an expression for the projected fractional overdensity for a survey of functional depth $L$ and uniform mean density $\rho$:
\begin{equation}
\Sigma(\vec R) = \frac{1}{L} \int dz\;\rho(\vec R, z)
\end{equation}\label{eq:wx}
\noindent where $\hat{z}$ is along the LOS and $\vec R$ is projected separation, as used in Section \ref{sec:alignment}. $L$ is a measure of how far along the LOS we average when measuring $\epsilon_{\textsc{LRG}}$. As our survey is not homogeneous, we generalize $L$ to an expression of $N(z)$. Using the weights we give each galaxy pair $w$, we sum over all combinations of color bins $B1$, $B2$, and galaxy pairs $j$, $k$. This is averaged per-bin and multiplied by the depth of that bin $B_d$. To account for clustering, we also much include the projected correlation function, $w_p$, as part of the bin depth\footnote{$w_p$ was estimated from DESI's Early Data Analysis using a $\Pi_\text{max}$ of $30 {\rm Mpc}$ (DESI Collaboration, in prep).}.
\begin{equation}
    L = (B_{\text{d}}+w_p)
    \frac{\Sigma_{B1}\Sigma_{B2}\Sigma_{\text{i}}\Sigma_{\text{j}}w(i, j)}{\Sigma_{B1}\Sigma_{\text{i}}\Sigma_{\text{i}}w(i, j)}
\label{eq:L_eq}
\end{equation}

\begin{figure}
\includegraphics[scale=.2]{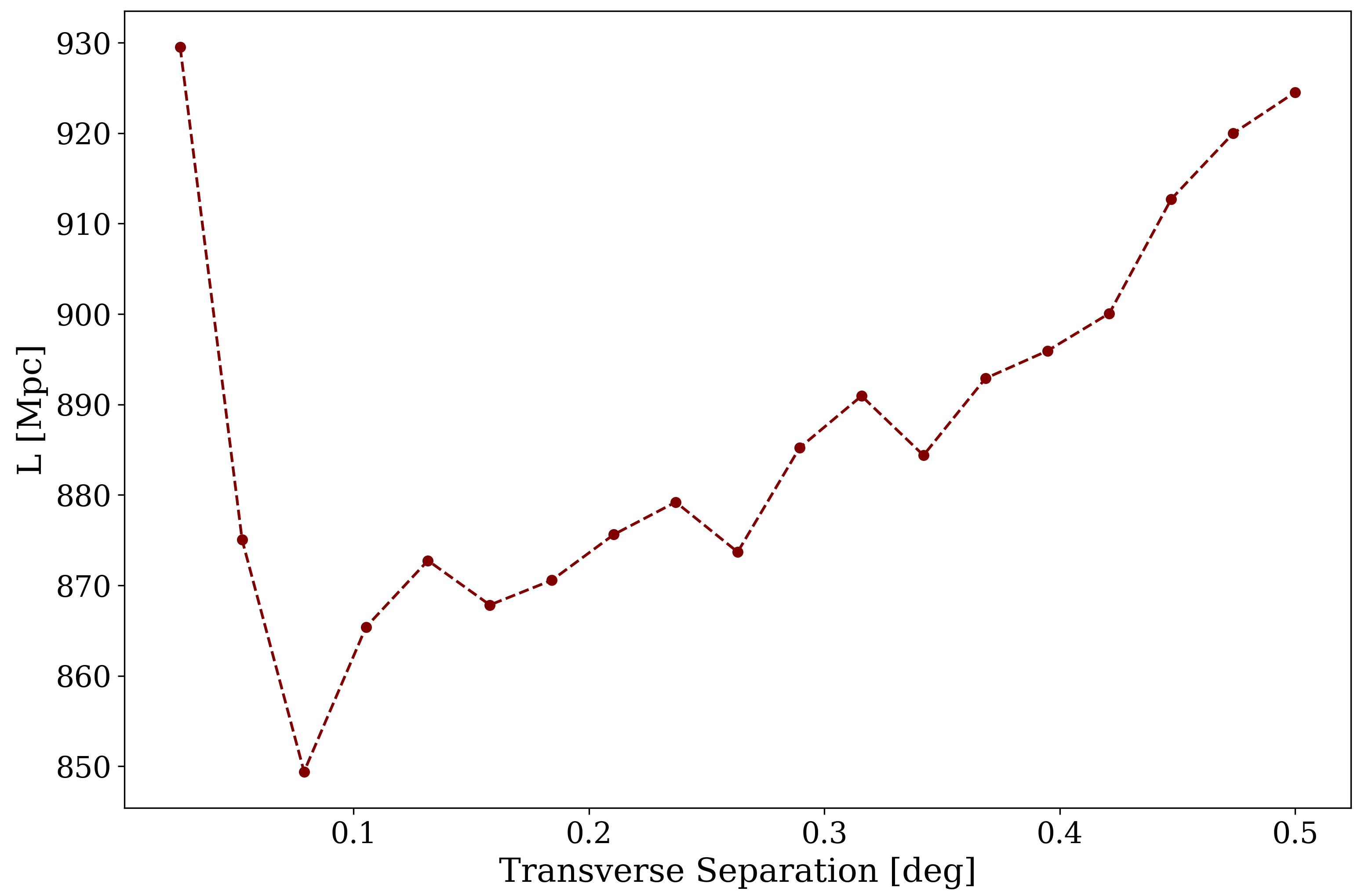}
\caption{$L(R)$, the effective radial distance that is averaged over when measuring $\mathcal{E}(R)$. This was estimated using Equation \ref{eq:L_eq} and a radial bin depth $B_d$ of $60 {\rm Mpc}$.}
\label{fig:L_est}
\end{figure}

We chose a $B_d$ of 60 ${\rm Mpc}$, which large enough to include enough pairs without averaging too far along the line of sight where our color weighting doesn't apply. This was calculated for each of the transverse $R$ bins used when measuring $\mathcal{E}(R)$, resulting in a function $L(R)$ (Figure \ref{fig:L_est}).\par 

The projected ellipticity is $\hat R_\alpha \epsilon_{\alpha \beta} \hat R_\beta$. For the average, we can just consider the $\hat R = \hat x$ direction. The shape-density correlation projected onto the plane of the sky is then given as
\begin{equation}
\mathcal{E}(R) = \left< \epsilon_{xx} \Sigma(R\hat x)\right> 
= -\frac{2\tau}{L}\left< (T_{yy}-T_{zz}) \int dz\;\rho(R\hat x, z)\right> 
\label{equation:shapedensity}
\end{equation}

\noindent As our model of $T$ is linear in the density field, it is straight-forward to compute this expectation value (Appendix \ref{appendix:wx}), yielding
\begin{equation}
\mathcal{E}(R) = \frac{\tau}{2L} R \frac{d}{dR}\left[\frac{1}{R} \Psi(R)\right]
\end{equation}
where we introduce
\begin{equation}
\Psi(R) = \int \frac{K\;dK}{2\pi} \frac{P(K)}{K} J_1(KR)
\end{equation}
$K$ is 2D Fourier Space, $P$ is the power spectrum, and $J_1$ is the first Bessel function.

$\tau$ can be inferred from our measurement of the shape-density correlation, $\left<\epsilon_{xx} \Sigma(R\hat x)\right>$, showing that the LOS shape and $\xi_2$ are correlated. We estimate $\tau$ as
\begin{equation}\label{eq:tau_obs}
\tau_{\rm obs} = \frac{2 L(R) \mathcal{E}(R)}{ R \frac{d}{dR}\left[\frac{1}{R} \Psi\right]},
\end{equation}
with our IA measurement, $\mathcal{E}$, and average over angular scales $R$. Writing this explicitly, if we measure $\mathcal{E}(R)$ from $R_0$ to $R_1$,
\begin{equation}
    \tau = \frac{1}{R_1-R_0} \int_{R_0}^{R_1} \tau_{\rm obs} dR
\end{equation}
This is our estimate of how the 3D ellipticity of galaxy shapes scales with the tidal field; it is directly proportional to the predicted $\xi_{\rm GI}$ signal.

\subsection{Shape - \texorpdfstring{$\xi_2$}{} Correlation}

Next, we turn to the correlation of shapes with the LOS $\xi_2$. To obtain the relation between 3D shapes and the LOS, we consider shapes viewed from a direction transverse to the LOS, i.e. an axis perpendicular to the projection axis above.

We define $\xi_2$ as $\xi(r,\mu) = \sum_\ell \xi_\ell(r) L_\ell(\mu)$,
with $\mu$ the cosine of the angle to the LOS. Therefore
the quadrupole signature $\xi_2$ is 
the correlation between the density at a point, here taken to be the origin, and the quadrupole-weighted density in spherical shells, $Q(r)$. This is
given as
\begin{equation}
Q(r) = 5\int \frac{d^2\hat r}{4\pi} \rho(\vec r) L_2(\mu)
\end{equation}
where $\int d^2\hat r$ indicates the 2D integral over the unit vector $\hat r$.

To express the transversely viewed shape, we take the average of $\epsilon_{xz}$ and $\epsilon_{yz}$, each
after the correction to a traceless quantity.  
For the $x$--$z$ projection, we have $\epsilon_{\alpha\beta} = \tau(T_{\alpha\beta}+T_{yy}/2)$, where the relevant quantity is $\epsilon_{zz}$.
Averaging with the $y$--$z$ projection, we have a transverse averaged eccentricity 
\begin{equation}
\epsilon_{zz} = \tau\left(T_{zz}+\frac{T_{xx}+T_{yy}}{4}\right) = \frac{3}{4}\tau T_{zz}
\end{equation}
Considering projections along $\hat x\pm \hat y$ also yield $T_{zz}/2$ as the 
only $m=0$ support.

The details of computing and simplifying $\left<\epsilon_{zz} Q(r)\right>$ can be found in Appendix \ref{appendix:eQ}, which result in
\begin{equation}\label{eq:eQ}
\left<\epsilon_{zz} Q(r)\right> = 
-\frac{\tau}{2}
\int \frac{q^2 dq}{2\pi^2} P(q) j_2(qr).
\end{equation}
This expression is averaged over radial bins of the correlation
function, resulting in averages of $j_2(qr)$. 

\subsection{Effect on Anisotropic Clustering \texorpdfstring{$\xi_2$}{}}
We expect the mean shape to be elongated along the LOS due to DESI's target selection, i.e. a non-zero mean $\epsilon_{zz}$ (Section \ref{section:polarization}). We call this LOS polarization $\epsilon_{\rm LRG}$. \par
Assuming $\epsilon_{zz}$ and the quadrupole signature $Q$ are Gaussian distributed, correlated, random variables, a non-zero $\left<\epsilon_{zz}\right>$ will result in a non-zero mean $Q(r)$ as
\begin{equation}
\left<Q\right> = \left<\epsilon_{zz}\right> 
\frac{\left<\epsilon_{zz}Q\right> }{\left<\epsilon_{zz}^2\right>}.
\end{equation}
where the expectation values come from summing over each galaxy. From our tidal model,
\begin{eqnarray}
\left<\epsilon_{zz}^2\right> &=& \bigg(\frac{3}{4}\tau\bigg)^2 \left<T_{zz}^2\right> \\
&=& \frac{\tau^2}{20} \int \frac{q^2 dq}{2\pi^2} P(q).
\end{eqnarray}

\noindent This integral is the variance in the density field $\sigma^2$,
hence $\left<\epsilon_{zz}^2\right> = \tau^2 \sigma^2/20$. We measured this as the variance in the shape parameter $\epsilon_1$ of all galaxies in the imaging survey.\par

Combining the above results, we obtain an expression for the quadrupole signature arising from GI alignment and a shape-dependent selection bias:
\begin{equation}\label{eq:xi_rsd}
\xi_{\rm GI} = \left<Q(r)\right> = \epsilon_{\rm LRG} \frac{\tau}{ 2\left<\epsilon_{zz}^2\right>}
\int \frac{q^2 dq}{2\pi^2} P(q) j_2(qr)
\end{equation}
A summary of the variables we measured for this estimate are listed in Table \ref{tab:variables}. $\xi_{\rm GI}$ depends linearly upon these values:
\begin{equation}
    \xi_{\rm GI} \propto \epsilon_{\rm LRG} \frac{\tau}{ \left<\epsilon_{zz}^2\right>} \propto \epsilon_{\rm LRG} \frac{L \mathcal{E}}{ \left<\epsilon_{zz}^2\right>}
\end{equation}

\section{Modeling DESI's selection effects}\label{section:polarization}
In Section \ref{sec:alignment} we measured how the shapes of galaxies projected onto the sky are aligned with the density field. To infer how this affects RSD measurements, we need to estimate the extent of DESI's orientation-dependent selection bias. Since pole-on galaxies have a higher surface brightness and are more likely to pass selection, we expect a net orientation of galaxies along the LOS, or polarization $\epsilon_{\textsc{LRG}}$. This is defined as the ellipticity (Equation \ref{eq:e1}) relative to the LOS.\par
We estimate this by using a parent catalog of LRGs which is similar to the sample described in Section \ref{sec:imaging}, except without the fiber magnitude cut. We assign each parent LRG a 3D galaxy light profile, then simulate images of each profile from all viewing angles, without any extinction from internal dust. The polarization is the average $\epsilon_{\text{1,LOS}}$ of all 3D profiles which pass selection.

\subsection{Parent Sample}\label{subsec:parent}
We estimate polarization using a subsection of DESI LRGs in an area of the sky with the best-resolved shapes, with right ascension and declination limits of 0h0m0s $<\alpha<$ 0h40m0s and 0$^{\circ}<\delta<$ 5$^{\circ}$. This is in the South Galactic Cap (SGC) and part of the Legacy Imaging Survey's DES region. This parent sample of 41120 objects has the same criteria as DESI's final target selection, except without the fiber z-magnitude cut of $z'_{\text{fiber}}<21.61$ for the SGC. The fiber magnitude comes from the light within a 1.\arcsec5 aperture after convolving the shape model with a standardized PSF. This somewhat isolates the fiber magnitude from seeing variations, so we can safely use shapes from an area with the best seeing without impacting the distribution of underlying shapes. As in Section \ref{sec:desi_catalog}, we also use shape parameters from the best-fit, non-circular, model. 

$0.95\%$ of this sample have the same fiber $z$-magnitude as total $z$-magnitude. This indicates that these objects are either stars or unresolved galaxies. We ignored these objects for our analysis, but a more thorough simulation would involve simulating galaxies through the T\textsc{ractor} pipeline, as is done with the Obiwan project \citep{kong_removing_2020}.

\subsection{Light Profiles}\label{sec:light_profiles}
Our light profile for each galaxy begin as a realization of 100,000 points. This representation allows us to rapidly apply the triaxial axis lengths, rotations, and projections, as well as to apply a 2D Gaussian PSF and the eventual fiber aperture cut.\par

The points for a given galaxy are distributed in 3D based on its best-fit shape model from the parent catalog. DESI's T\textsc{ractor} pipeline represents projected galaxy shapes as a mixture of Gaussians \citep{hogg_replacing_2013}. To de-project these into 3D profiles, we take advantage of the fact that a 3D Gaussian projects to a 2D Gaussian. Therefore the 2D Gaussian mixture fits allow us to immediately construct a 3D model. This was done for all parent LRGs with a best-fit profile of de Vaucouleurs, exponential, and round-exponential LRGs. Relatively few LRGs were fit best with a Sersic profile. These tend to be bright enough that they are not near the aperture magnitude cut and therefore less affected by this biased selection; for simplicity we modeled these with a Hernquist profile \citep{hernquist_analytical_1990}. \par
\begin{figure}
\includegraphics[scale=.45]{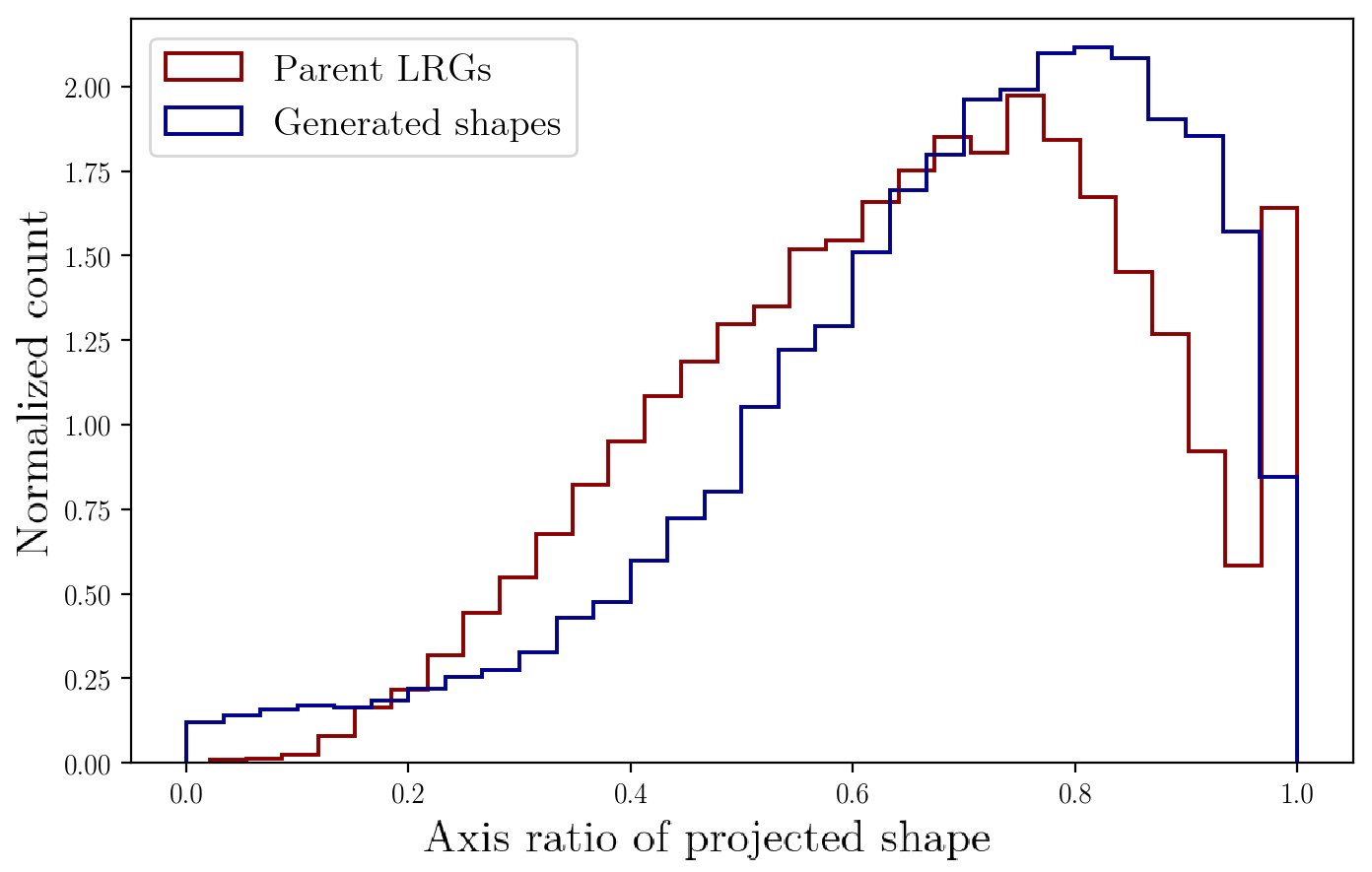}
\caption{A comparison of the axis ratios of LRGs in our parent sample and the projected axis ratios from a distribution of triaxial shapes. These are the triaxial shapes used in our polarization estimate. The spike at $b/a=1$ in the parent sample is artificial, likely due to poor shape fitting.}
\label{fig:sim_shapes}
\end{figure}

\begin{figure*}
\includegraphics[scale=.27]{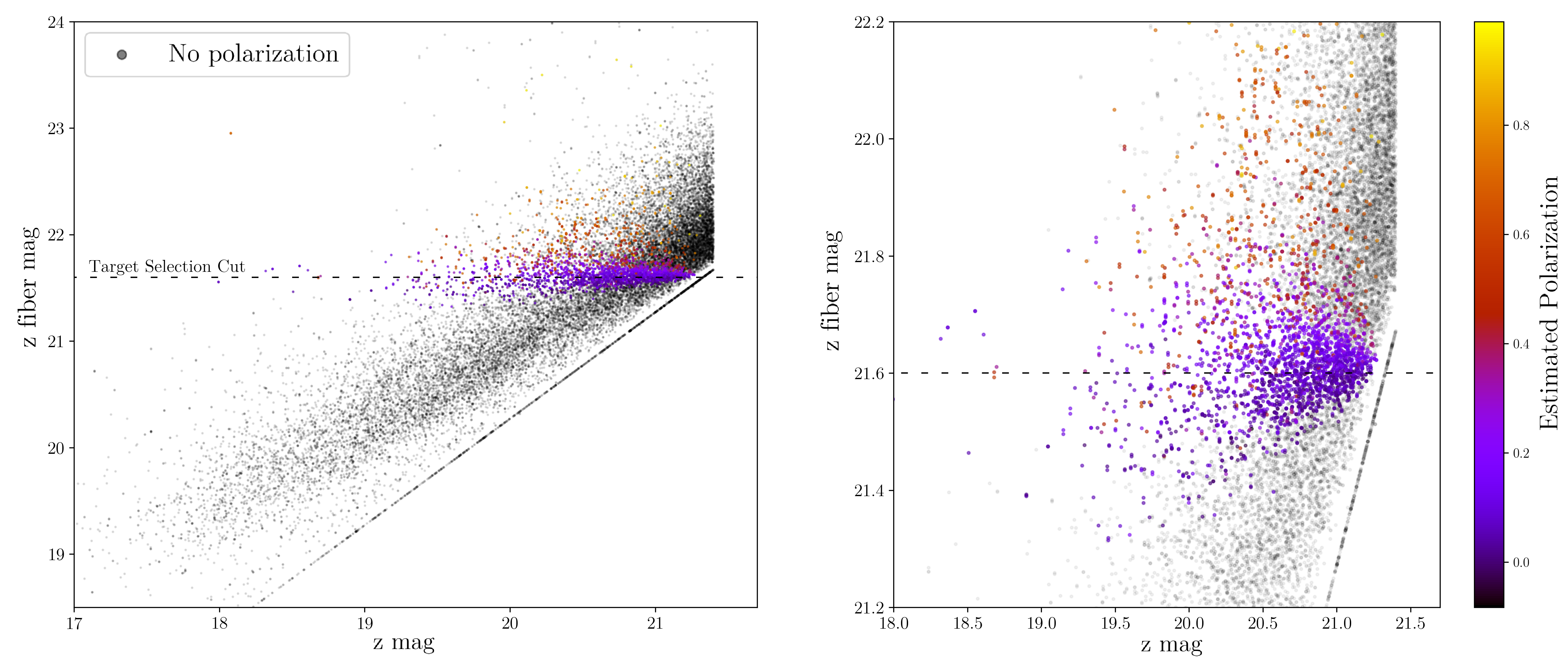}
\caption{Results from our N-body reproduction of DESI's target selection. There are two main flux cuts on the LRGs: a sliding $r-W1$ vs $W1$ cut which dominates at bluer colors, and the $z_{\text{fiber}}$ cut which dominates at redder colors. The full parent sample is shown on the left and a closer look near the fiber magnitude cut on the right. Each galaxy was assigned a triaxial shape, which was rotated to 100 random orientations. Its polarization is the average ellipticity relative to the light of sight of the objects which passed an aperture-magnitude cut. For target selection, we find that the orientations of shapes matters only for objects very close to the fiber magnitude cut, and is more likely to matter for more elliptical galaxies.}
\label{fig:polarization}
\end{figure*}

\begin{figure*}
\subfloat[]{\includegraphics[scale=.32]{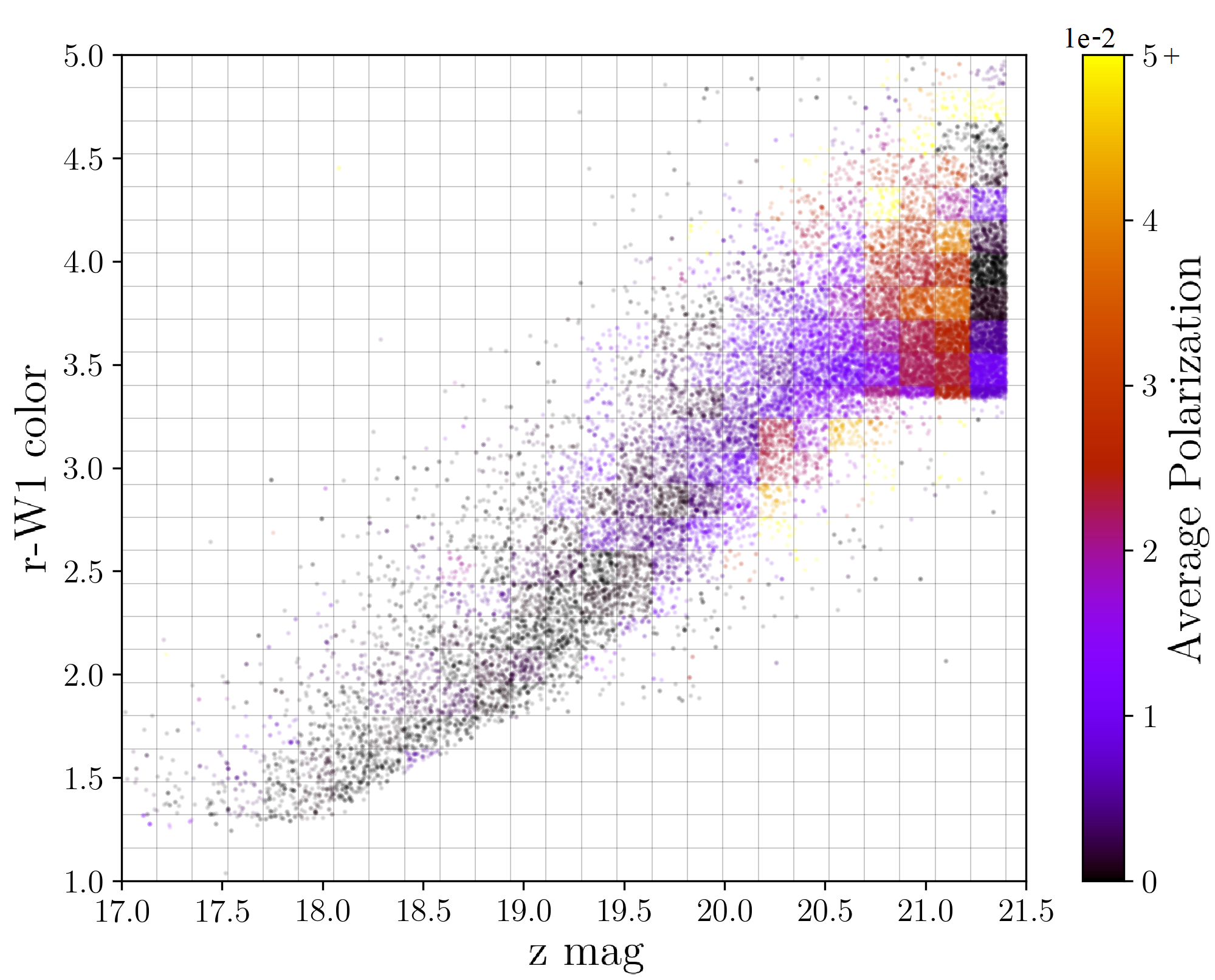}}
\hfill
\subfloat[]{\includegraphics[scale=.38]{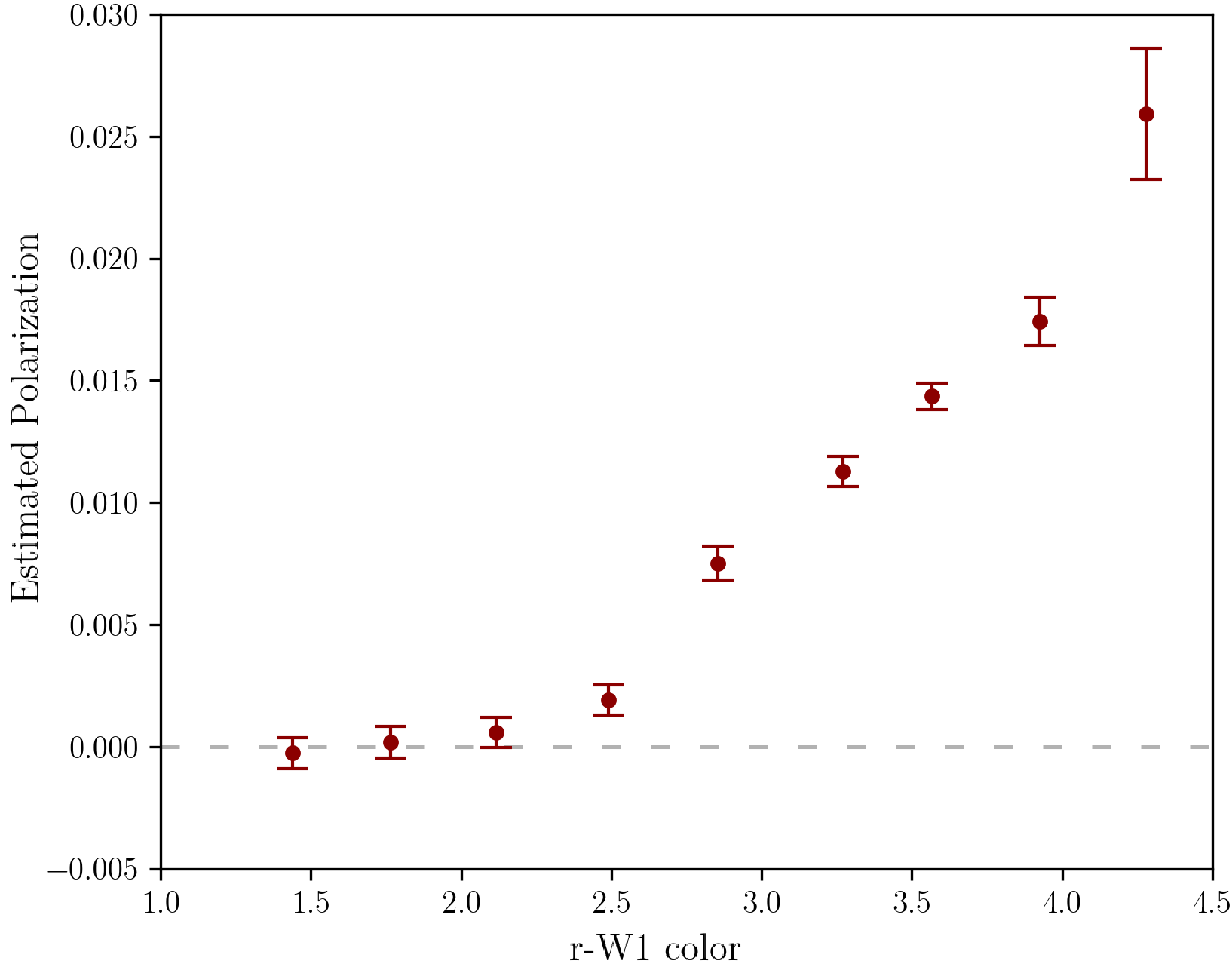}}
\caption{Properties of selected targets from our simulated images. a: the $z$ mag and color taken from a parent sample of LRGs. Each square is colored by the average polarization in that bin. We attribute the drop in polarization near the highest $z$ mag to a selection effect: in order for these targets to pass selection they must have a fiber magnitude very close to their total magnitude, resulting in more compact shapes and a dampened polarization. b: the polarization of selected targets binned by color. We expect this trend, since fainter galaxies tend to fall closer to the fiber-magnitude cut (Figure \ref{fig:polarization}). A higher polarization for redder colors could lead to an increased $\xi_2$ bias at higher redshifts and mimic structure growth.}
\label{fig:polarization_selected}
\end{figure*}

\subsection{Polarization Estimate}\label{sec:polarization}

For each object in the parent sample, we assign a triaxial shape based on its projected shape. These were randomly drawn from the expected distribution of triaxial shapes for bright ($r$-band absolute magnitudes $>-19$), medium ($2 < r$-band radius $ < 7 h^{-1}kpc$) ellipticals in imaging from the Sloan Digital Sky Survey \citep{padilla_shapes_2008}. 41120 3D shapes were projected along a random viewing angle and ranked by the axis ratio of the resulting ellipse. The LRG parent sample was also sorted by axis ratio, and matched with the triaxial shape corresponding to the projected shape of the same rank.\par

To test these triaxial shapes, we viewed them each from a different angle and compared the projected axis ratios to our parent sample (Figure \ref{fig:sim_shapes}). These distributions are not identical; note the artificial spike in the parent sample at $b/a=1$ which is likely from poor shape fitting. Differences in the distributions could also be due to shape-dependent fitting biases in T\textsc{ractor}, or imperfect distributions from \cite{padilla_shapes_2008}, including shape evolution from z $=0$ or internal obscuration.

The point positions from Section \ref{sec:light_profiles} were scaled by the assigned three axis lengths for each galaxy. They were then rotated to 100 random orientations and projected along one axis. The resulting `images' were scaled using the ratio of the observed half-light radius and the average half-light radius of all model orientations. We next need to emulate an observation in  1\arcsec seeing. Instead of convolving with a Gaussian, we took the quicker approach of adding pre-computed, 2D deflections to the projected points. The fiber magnitude was estimated by from the fraction of points which fell within an 1.\arcsec5 - diameter aperture, and the observed total magnitude of the LRG. The light profiles used did not perfectly replicate the observed $z_{\text{fiber}}$ values, so we added a calibration factor to the N-body fiber magnitude for each of the four light profiles to match the true $z_{\text{fiber}}$ median. Objects with a fiber magnitude less than 21.61 passed selection.\par

For each simulated image which passed selection, we measured the corresponding 3D profile's complex ellipticity relative to the LOS. This is the same convention as Equation \ref{eq:e1}, except shapes are projected in the transverse direction. The average of these is our polarization $\epsilon_{\rm LRG}$. 54.2\% of our simulated galaxy images passed the fiber magnitude cut, similar to the actual value of 52.9\%. The polarization for these galaxies is $0.0087\pm0.0002$. By determining the selection of a set of orientations for each galaxy shape, we can also estimate which galaxies in the original sample may have an orientation-dependent selection (Figure \ref{fig:polarization}). To see what polarization DESI can expect in its targets, we've plotted the average polarization in bins of $z$ mag and $r-W1$ color (Figure \ref{fig:polarization_selected}a). \par
We also find that the redder LRGs may be more affected by orientation. This translates to a correlation between redshift and polarization, which could affect studies of structure evolution (Figure \ref{fig:polarization_selected}b).

\begin{figure}
\includegraphics[scale=.45]{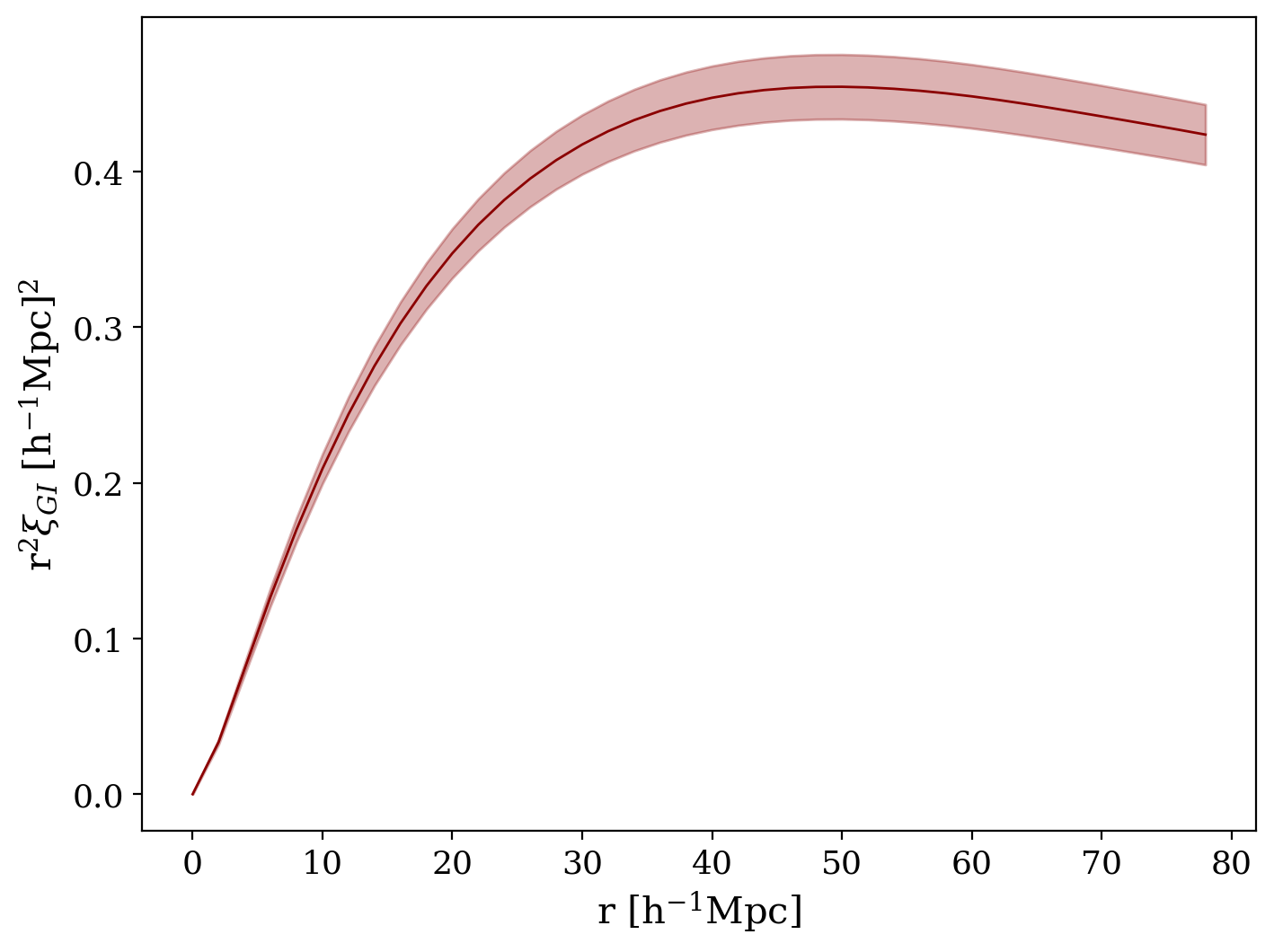}
\caption{Estimated impact on DESI's measurement of the RSD quadrupole. This about corresponds to a 0.5\% error for separations around 40-80 $h^{-1} {\rm Mpc}$ based on SDSS-III BOSS measurements \citep{anderson_clustering_2014}. The precision of our estimate is shown in the shaded region as the standard error.}
\label{fig:xi_rsd}
\end{figure}

\section{Estimate of False RSD signature \texorpdfstring{$\xi_{\textsc{GI}}$}{} in DESI}\label{sec:result}
At this point, we have measured all the necessary components to estimate the $\xi_2$ signature arising from IA and DESI's selection bias. A summary of the variables used in this estimate are listed in Table \ref{tab:variables}.\par

$\epsilon_{\rm LRG}$, the polarization of galaxy shapes along the LOS, is measured in Section \ref{section:polarization}. $\langle \epsilon_{zz}^2\rangle$ is the variance of the real part of the complex ellipticities which describe the shapes of DESI's LRGs and is 0.031. We used the power spectrum, $P(k)$ from A\textsc{bacus}S\textsc{ummit} \citep{maksimova_abacussummit_2021}.\par

$\tau$ is a function of effective depth $L$, or how far along the line of sight we average when measuring $\epsilon_{\textsc{LRG}}$. This was estimated using the color weighting scheme from Section \ref{sec:weighting}, has a value around $L=620 h^{-1}{\rm Mpc}$, and can be seen in Figure \ref{fig:L_est}. $\tau$ also depends on the projected shape-density correlation of LRGs $\mathcal{E}(r)$ which we measured in Section \ref{sec:alignment}. Averaging over the bins of projected separation, we estimate $\tau_{\rm obs}=-0.131$.\par

Using equations \ref{eq:xi_rsd} and \ref{eq:tau_obs} to bring everything together, we determine $r^2\xi_{\rm GI}$ to be 0.41($h^{-1}$Mpc)$^2$ around 10-80 $h^{-1}{\rm Mpc}$. The full separation dependence is shown in Figure \ref{fig:xi_rsd}. SDSS-III measures $r^2\xi_2$ at these scales to be near 75 ($h^{-1}$Mpc)$^2$ \citep{anderson_clustering_2014}. This puts our estimate of the fractional error on $\xi_2$ around 0.5\% at 40-80 $h^{-1}$Mpc.

\input{derivation_details/variable_guide.tex}

\section{\texorpdfstring{$\xi_\textsc{GI}$}{} Estimate in Abacus}
To demonstrate that an aperture selection produces a $\xi_2$ signature and test our linear tidal model connecting the GI and RSD signals, we next model the problem using A\textsc{bacus} S\textsc{ummit} simulations.

As in Section \ref{sec:abacus}, we started with a 2000 $h^{-1}$Mpc box of large halos and mapped their positions to redshift, right ascention, and declination. Sky cuts were applied to ensure a uniform sky distribution at each redshift. 3D Sersic profiles of 100,000 points were generated for each halo, as in Sections \ref{sec:light_profiles}-\ref{sec:polarization}, except using the halo's original triaxial shape. The half-light radius used for each halo was drawn from a distribution matching the physical radii of the DESI LRG parent sample and scaled using the average half-light radii of the point profile projected to 10 random orientations. We counted the number of points which fell within a 1.\arcsec5 aperture and measured the shape of each halo projected both on the sky and relative to the LOS.

To see how an aperture selection impacts the $\xi_2$ measurement, we created two samples: one without any selection, and one only with halos containing more than than 48,000 points within the aperture, which corresponds to 50\% of the halos. We measured $\xi_2(r)$ for both sets in real space space and in redshift space, using the halo's original velocities.\par
\begin{figure}
\includegraphics[scale=.39]{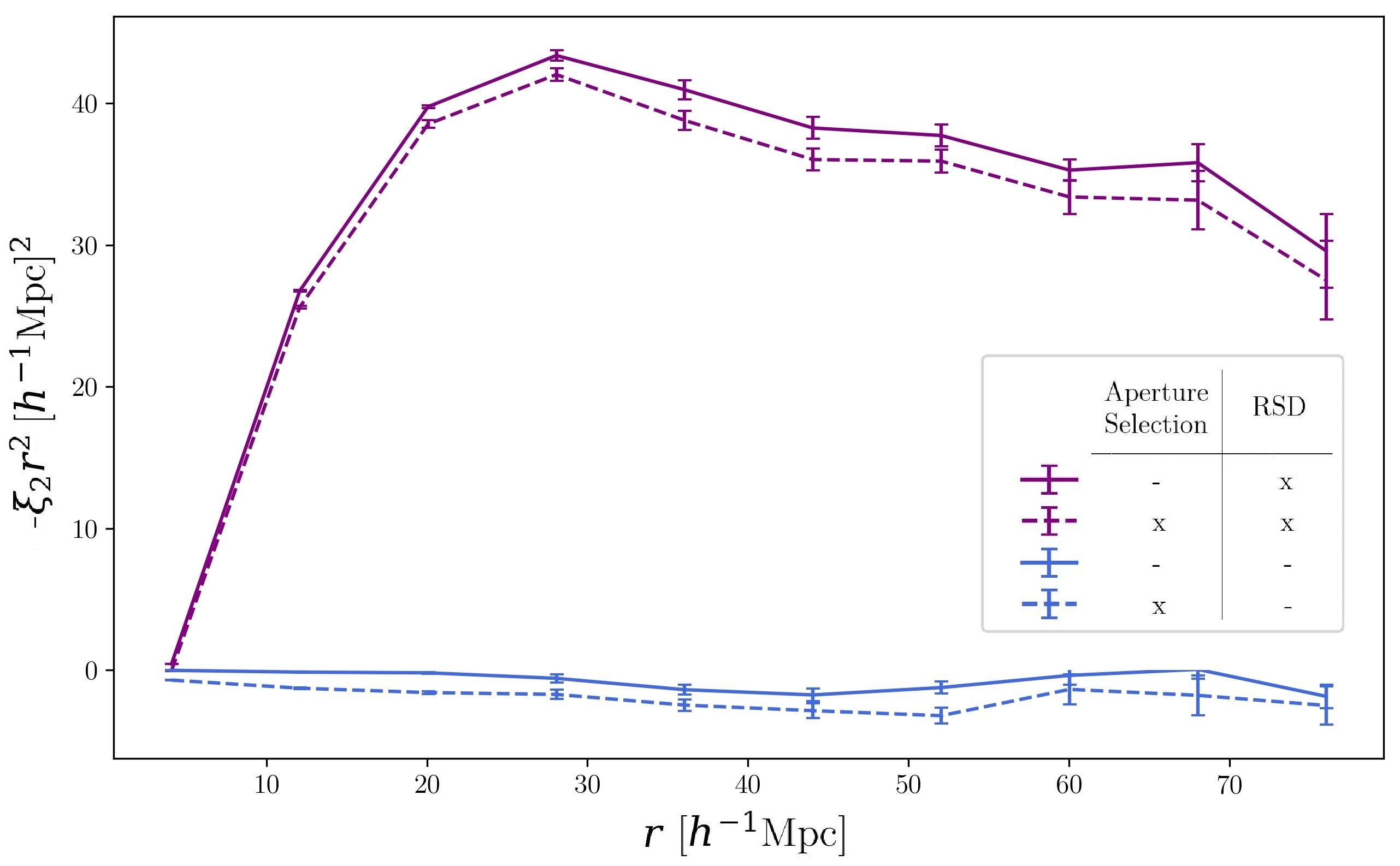}
\caption{$\xi_2$ measurements of A\textsc{bacus} halos catalogs with and without an aperture-based selection. The top two lines include Redshift-Space Distortions. The aperture-based selection creates an artificial, non-zero RSD signature, which acts in opposition to $\xi_2$ on large scales.}
\label{fig:xi_measures}
\end{figure}

$\xi_2(r)$ was determined using the Landy-Szalay Estimator \citep{landy_bias_1993} and averaged over 10 sets of randoms, generated with random right ascension and declinations for each redshift. This entire process was done for 5 A\textsc{bacus} S\textsc{ummit} simulation boxes, and their average $\xi_2(r)$ and standard error is shown in Figure \ref{fig:xi_measures}. 

\begin{figure}
\includegraphics[scale=.44]{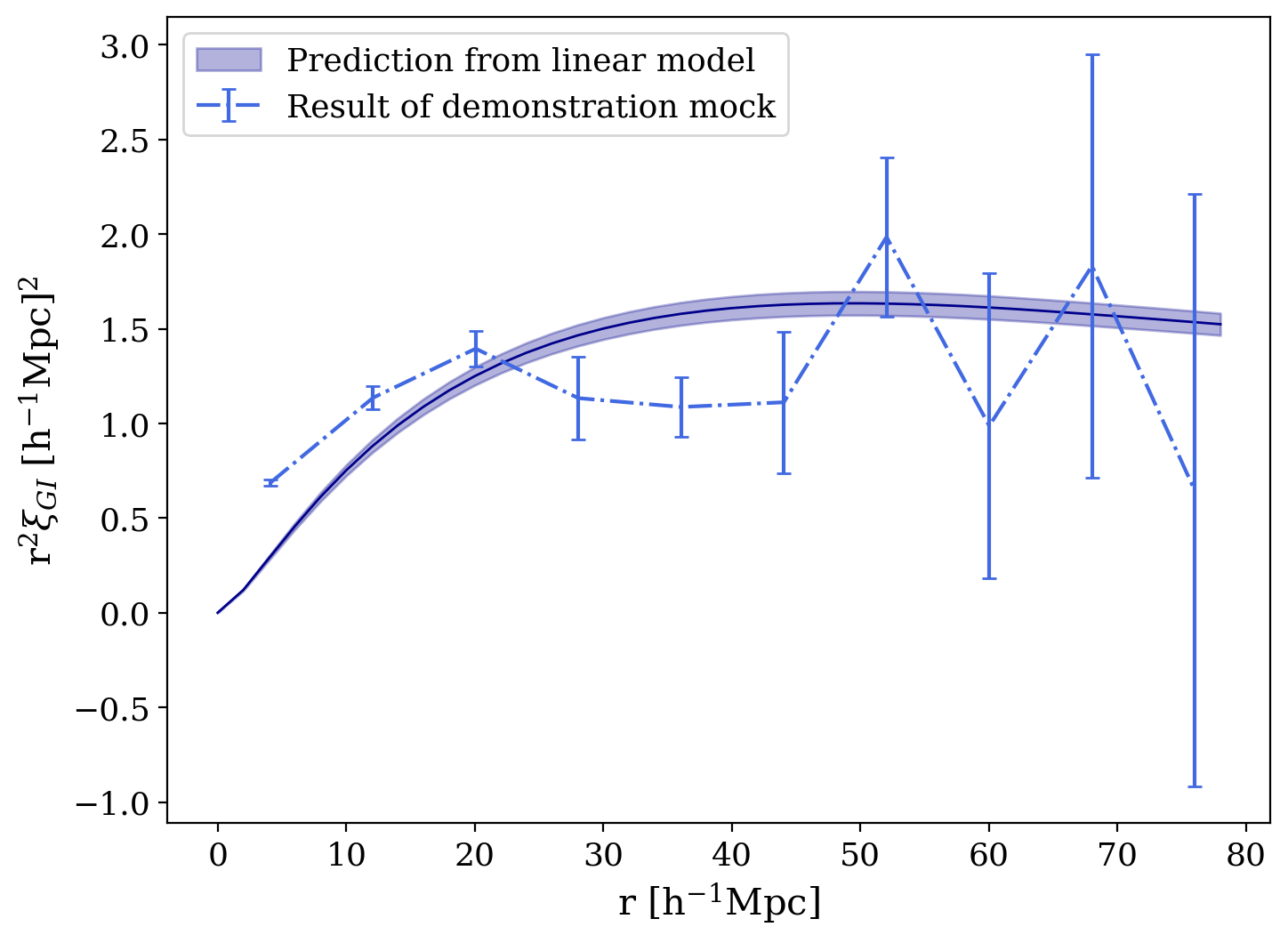}
\caption{The artificial RSD signature induced by an aperture-based selection. Here we compare the prediction made with our linear tidal model to the measured difference in $\xi_2$ between a halo catalog with and without the aperture selection. This is the difference between the measurements of $\xi_r$ in Figure \ref{fig:xi_measures} without RSD.} 
\label{fig:xi_ab}
\end{figure}

As in Section \ref{sec:result}, we used our linear tidal model to predict the $\xi_2$ bias caused by the aperture selection for this halo catalog. We measured the projected intrinsic alignment of the halo catalog in radial bins which resulted in an average survey depth, $L$, of around 580 $h^{-1}$Mpc between $0.1-0.5$ deg. The polarization due to aperture cut was $\epsilon_{\textsc{LRG}}=7.6\pm0.1\times 10^{-3}$. The resulting prediction is compared to the model in Figure \ref{fig:xi_ab}.

We expect the bulk of the disagreement between these two simple models to be due to the linear approximation, which does not hold at lower separations, and simplifications in the demonstration mock. The largest simplification here is that every galaxy is modeled with a Hernquist light profile. Any profiles which are denser than reality will underestimate the polarization due to aperture selection. However, the A\textsc{bacus} approximation is comparable to the prediction from the linear model and serves as a adequate demonstration of how a false $\xi_2$ signature can arise for DESI.

\section{Conclusion}
The objective of this study is to determine the approximate impact on DESI's RSD measurements due to an orientation bias in LRGs. We have demonstrated that the effect is significant for DESI and estimate a 0.5\% fractional decrease of $\xi_2$ for separations of 40-80 $h^{-1} {\rm Mpc}$. DESI forecasts a total $f\sigma_8$ around 0.4-0.7\% (with ELG and LRGs combined), so it is important to mitigate this effect. 

To reduce the effects of intrinsic alignment for DESI, simple yet severe choices involve only measuring $\xi_2$ in galaxy subsamples, perhaps cut by total magnitude or color. More practically, our estimate could be used for calibration.

As the DESI survey progresses and the precision in $\xi_2$ increases, there are several opportunities to improve our bias estimate. Our estimate is directly proportional to the measured polarization $\epsilon_{\textsc{LRG}}$ and IA signal $w_{\text{x}}$, both of which include systematic uncertainties.
The main systematic uncertainty in our polarization estimate arises from the choice of the triaxial shape distribution. We expect the majority of our galaxies to be prolate \citep{padilla_shapes_2008}, which are more affected by selection bias than oblate and result in a higher polarization. We match the expected distribution of projected shapes in a region of the sky with the best shape fits, but 5.6\% of galaxies in this subsample are fit as circles, creating an artificial spike at $b/a=1$ (Figure \ref{fig:sim_shapes}). A better estimate could be made with more accurate shapes, ie from the Dark Energy Survey \citep{gatti_dark_2021}, or reproducing \cite{padilla_shapes_2008} with DESI's LRGs.\par
Although partially mitigated by color weighting, the IA signal in this work is reduced by weak lensing and diluted by the inclusion of pairs which have large radial separations. We also expect a 5-10\% uncertainly in our forecast due to the difficulty in accurately estimating $L$ with photometric distances. This will be drastically improved with DESI's first year of data, which contains 2.5 million quality LRG spectra. The LOS distance we average over due to uncertainty in radial distances, $L=865 h^{-1}{\rm Mpc}$, will decrease by a factor of at least 20 with redshifts. Advancing our ability to measure IA for only pairs of galaxies which are physically associated will be the strongest improvement to the false $\xi_{2}$ estimate. 

\section*{Acknowledgements}

CL wishes to acknowledge Sean Moss for his support, software and otherwise, and Tanveer Karim for many helpful conversations. The authors also thank Thomas Bakx for his valuable feedback and Eric Jullo, Eusebio Sanchez, Robert Kehoe, and Zachary Slepian for their input through the collaboration review process.\par
This material is based upon work supported by the National Science Foundation Graduate Research Fellowship under Grant No. DGE1745303, the U.S.\ Department of Energy under grant DE-SC0013718, and NASA under ROSES grant 12-EUCLID12-0004.  DJE is further supported by the Simons Foundation.
\par
This research is also supported by the Director, Office of Science, Office of High Energy Physics of the U.S. Department of Energy under Contract No. DE–AC02–05CH11231, and by the National Energy Research Scientific Computing Center, a DOE Office of Science User Facility under the same contract; additional support for DESI is provided by the U.S. National Science Foundation, Division of Astronomical Sciences under Contract No. AST-0950945 to the NSF’s National Optical-Infrared Astronomy Research Laboratory; the Science and Technologies Facilities Council of the United Kingdom; the Gordon and Betty Moore Foundation; the Heising-Simons Foundation; the French Alternative Energies and Atomic Energy Commission (CEA); the National Council of Science and Technology of Mexico (CONACYT); the Ministry of Science and Innovation of Spain (MICINN), and by the DESI Member Institutions: \url{https://www.desi.lbl.gov/collaborating-institutions}.\par

The DESI Legacy Imaging Surveys consist of three individual and complementary projects: the Dark Energy Camera Legacy Survey (DECaLS), the Beijing-Arizona Sky Survey (BASS), and the Mayall z-band Legacy Survey (MzLS). DECaLS, BASS and MzLS together include data obtained, respectively, at the Blanco telescope, Cerro Tololo Inter-American Observatory, NSF’s NOIRLab; the Bok telescope, Steward Observatory, University of Arizona; and the Mayall telescope, Kitt Peak National Observatory, NOIRLab. NOIRLab is operated by the Association of Universities for Research in Astronomy (AURA) under a cooperative agreement with the National Science Foundation. Pipeline processing and analyses of the data were supported by NOIRLab and the Lawrence Berkeley National Laboratory. Legacy Surveys also uses data products from the Near-Earth Object Wide-field Infrared Survey Explorer (NEOWISE), a project of the Jet Propulsion Laboratory/California Institute of Technology, funded by the National Aeronautics and Space Administration. Legacy Surveys was supported by: the Director, Office of Science, Office of High Energy Physics of the U.S. Department of Energy; the National Energy Research Scientific Computing Center, a DOE Office of Science User Facility; the U.S. National Science Foundation, Division of Astronomical Sciences; the National Astronomical Observatories of China, the Chinese Academy of Sciences and the Chinese National Natural Science Foundation. LBNL is managed by the Regents of the University of California under contract to the U.S. Department of Energy. The complete acknowledgments can be found at \url{https://www.legacysurvey.org/}.\par

The authors are honored to be permitted to conduct scientific research on Iolkam Du’ag (Kitt Peak), a mountain with particular significance to the Tohono O’odham Nation."

\section*{Data Availability}
The DESI Legacy Imaging Survey is publicly available at \href{https://www.legacysurvey.org/}{legacysurvey.org}. AbacusSummit simulations are publicly available at \href{https://abacusnbody.org/}{abacusnbody.org}. Code for projecting ellipsoids and generating light profiles can be found at \href{https://github.com/cmlamman/ellipse_alignment}{github.com/cmlamman/ellipse\_alignment}.\par
All data plotted in this paper are available at \href{https://doi.org/10.5281/zenodo.7058448}{zenodo.org/record/7058448}



\bibliographystyle{mnras}
\bibliography{references} 



\onecolumn
\appendix

\section{Projection of Triaxial Ellipsoids}\label{appendix:projection}
\input{derivation_details/projecting_ellipsoids.tex}

\section{Expanded Derivations}

\subsection{Weak Lensing Estimate}\label{appendix:wl}
\input{derivation_details/lensing_derivation.tex}

\subsection{Shape-Density Correlation}\label{appendix:wx}
\input{derivation_details/derivation_1.tex}

\subsection{Shape - \texorpdfstring{$\xi_2$}{} Correlation}\label{appendix:eQ}
\input{derivation_details/derivation_2.tex}


\bsp	
\label{lastpage}
\end{document}

%% file: derivation_details/variable_guide.tex
\begin{table*}
    \centering
    \caption{A summary of the variables used in estimating the quadrupole signature arising from intrinsic alignment, and their measured values in DESI's LRGs sample.}\label{tab:variables}
    \begin{tabular}{llc} 
    	\hline
    	Variable & Description & Measured Value\\
        	\hline
        	$R$ & Projected separation on plane of the sky & -\\
        	$\mathcal{E}(R)$ & Intrinsic alignment, i.e. mean ellipticity of one galaxy relative to the projected separation to another& Figure \ref{fig:ia_main}\\
        	$L(R)$ & Depth of measurement along LOS when measuring $\mathcal{E}(R)$& Figure \ref{fig:L_est}\\
        	$\epsilon_{\textsc{LRG}}$ & Polarization of LRG shapes along the LOS ($+\hat{z}$), equivalent to $\langle\epsilon_{zz}\rangle$ & $7.6\pm0.1\times 10^{-3}$\\
        	$\langle \epsilon_{zz}^2\rangle$ & Averaged ellipticity relative to the LOS, measured as variance in the real part of $\epsilon_1$ & 0.031\\
        	$\tau$ & How 3D ellipticity scales with the tidal tensor, galaxy axis lengths behave as $I+\tau T_{ij}$ & -0.131\\
        	$\xi_{\textsc{GI}}(r)$ & Quadrupole signature arising from intrinsic (GI) alignment & Figure \ref{fig:xi_rsd}\\
        	\hline
    \end{tabular}
\end{table*}


%% file: derivation_details/projecting_ellipsoids.tex
This section details how we obtained the axis ratios and orientations of projected triaxial ellipsoids for our mock catalogs. We adapted the method derived in \cite{gendzwill_analysis_1981} to project ellipsoids onto the celestial sphere.\par

We denote the 3 ellipsoidal axis lengths as $\lambda_j$, $j=1,2,3$.  We then define the diagonal matrix ${\bf\Gamma}$ such that $\Gamma_{ij} = \delta_{ij}\lambda_j^{-2}$, where $\delta$ is a Kronecker delta.  We normalize the corresponding axis directions $\vec{s}_j$ and organize them as rows of a matrix ${\bf S}$, so that $S_{ij}$ is the $j^{\rm th}$ component of the $i^{\rm th}$ vector.  
We are projecting along the $\hat x$ unit vector direction, here denoted as component 1, onto the $\hat y-\hat z$ plane.

We define the column vector $\vec{m}$ as 
\begin{eqnarray}
\vec{m} = \left(\hat x^T {\bf S}^T{\bf\Gamma S} \hat x\right)^{-1} \hat x^T {\bf S}^T{\bf\Gamma S}
\end{eqnarray}
where the pre-factor adopts the normalization that $\vec{m}\cdot\hat{x}=1$.
We then compute vectors 
$\vec{u}$ and $\vec{v}$
with elements 
$u_j = \hat{y}\cdot(\vec{m} \times \vec{s}_j)$ and
$v_j = \hat{z}\cdot(\vec{m} \times \vec{s}_j)$, written alternatively as
\begin{eqnarray}
u_j &=& m_1 S_{j3} - m_3 S_{j1} \\
v_j &=& m_1 S_{j2} - m_2 S_{j1}.
\end{eqnarray}
We use these to compute the scalars
$A = \vec{u}^T {\bf \Gamma} \vec{u}$, 
$B = \vec{u}^T {\bf \Gamma} \vec{v}$, and
$C = \vec{v}^T {\bf \Gamma} \vec{v}$.
\\
The orientation angle of the projected ellipse's primary axis, measured in the $+\hat{y}$ direction from $\hat{z}$ is
\begin{eqnarray}
    \tan 2\theta = \frac{-2B}{A-C}
\end{eqnarray}
\\
And the minor and major axis lengths of the ellipse, $b$ and $a$ are given as:
\begin{eqnarray}
        &\frac{1}{a^2} = \frac{A+C}{2} + \frac{A-C}{2\cos{2\theta}} \\
        &\frac{1}{b^2} = A+C - \frac{1}{a^2}
\end{eqnarray}
To project the shapes on the sky, we rotated the original ellipsoid eigenvectors using the object's right ascension and declination, so that $\hat{x}$ lay along the LOS. This results in the axis lengths and orientation angle measured East of North for each halo. A function which performs these operations is available here: \href{https://github.com/cmlamman/ellipse_alignment}{github.com/cmlamman/ellipse\_alignment}\par

%% file: derivation_details/lensing_derivation.tex
Here are the details of how we obtained the expressions of surface over density (Equations \ref{eq:lens_1} and \ref{eq:lens_2}) used in the lensing estimation. $r_0 = 7.78\ {\rm Mpc}/ h$ is the 3D correlation length for DESI clustering \citep{kitanidis_imaging_2020}, $\beta = 2.15$ is the clustering bias for DESI LRGs \citep{zhou_clustering_2021}, and $\rho_0 = 2.68\times 10 ^{-30} \text{ g }\text{cm}^{-3}$ is the critical matter density of the Universe from Planck 2018 \citep{collaboration_planck_2020}. We start with an expression for the surface overdensity at a projected separation $r_p$:
\begin{eqnarray}
    \Sigma (r_p) = \int_{-\infty}^{+\infty} \rho_0 \xi_{gm}dz
\end{eqnarray}
where we assume the density follows $\langle\rho_m(r)\rangle = \rho_0\xi_{gm}$. For the correlation function, we assume a power-law model $\xi_{gg} = (r_0/r)^2$ where $\xi_{gm} = \frac{1}{\beta}\xi_{gg}$. Therefore the projected correlation function can be expressed as:
\begin{eqnarray}
    w_p(r_p) = \int\xi_{gg} dz = \int_{-\infty}^{+\infty}\frac{r_0^2}{r_p^2 + z^2}dz
    = \pi\frac{r_0^2}{r_p}
\end{eqnarray}
The surface overdensity becomes
\begin{eqnarray}
    \Sigma (r_p) = \frac{\rho_0}{\beta}w_p(r_p) = \pi\frac{\rho_0}{\beta}\frac{r_0^2}{r_p}
\end{eqnarray}
We integrate this over $r_p'$ to get an expression for the average surface overdensity within $r_p$:
\begin{eqnarray}
    \bar{\Sigma} (<r_p) = \frac{1}{\pi r_p} \int_0^{r_p^2} \Sigma(r_p')2\pi r_p'dr_p' = 
    2\pi\frac{\rho_0}{\beta}\frac{r_0^2}{r_p}
\end{eqnarray}

%% file: derivation_details/derivation_1.tex
Starting from Equation \ref{equation:shapedensity}, we can continue the computation as:
\begin{eqnarray}
\mathcal{E}(R) &=& \frac{\tau}{2L} \int dz\ \int \frac{d^3 q}{(2\pi)^3} 
\left[\frac{q^2_y - q^2_x }{ q^2} \right] \left.e^{-i\vec q\cdot \vec x}\right|_{\vec x= 0} 
\int \frac{d^3k}{(2\pi)^3} \left.e^{i\vec k\cdot \vec r}\right|_{\vec r = (R,0,z)} \left<\tilde\rho^*(\vec q) \tilde\rho(\vec k)\right> \\
&=& \frac{\tau}{2L}\int dz\ \int \frac{d^3k}{ (2\pi)^3}\left[k^2_y-k^2_x\right] k^{-2} P(k) \left.e^{i\vec k\cdot \vec r}\right|_{\vec r = (R,0,z)}.
\end{eqnarray}

Next, the integral over $z$ creates $\int dz\ \exp(i k_z z) = (2\pi) \delta^D(k_z)$.  We denote the space of $(k_x,k_y)$ as $\vec K$, and similarly $\vec R$ as $(x,y)$.  
So we have
\begin{equation}
\mathcal{E}(R) = \frac{\tau}{2L} \int \frac{d^2 K}{(2\pi)^2} \left(K_y^2 - K_x^2\right) K^{-2} P(K) e^{iK_x R}.
\end{equation}
To simplify this, we introduce 
\begin{equation}
\Phi(\vec R) = \int \frac{d^2 K}{(2\pi)^2} \frac{P(K)}{ K^2} e^{i\vec K\cdot\vec R},
\end{equation}
which in turn implies
\begin{equation}
\mathcal{E}(R) = \frac{\tau}{2L} \left(\partial^2_y - \partial^2_x\right)\left.\Phi(\vec R)\right|_{\vec R = R\hat x}.
\end{equation}
$\Phi(\vec R)$ is isotropic, and can be simplified to a Hankel transform
\begin{equation}
\Phi(R) = \int \frac{K\ dK}{2\pi} \frac{P(K)}{ K^2} J_0(KR)
\end{equation}
with $J_0$ being the Bessel function.
For a general function $f(R)$, we have
$\partial^2 f/\partial x^2 = \partial^2 f/\partial R^2$ 
and 
$\partial^2 f/\partial y^2 = (1/R) \partial f/\partial R$.
So we have 
\begin{equation}
\mathcal{E}(R) = \frac{\tau}{2L} \left(\frac{1}{R}\partial_R - \partial^2_R\right)\Phi(R) = \frac{\tau}{2L} R \frac{d}{dR}\left[\frac{1}{R} \Psi(R)\right]
\end{equation}
where we introduce
\begin{equation}
\Psi(R) = -\frac{d\Phi}{dR} 
= \int \frac{K\;dK}{2\pi} \frac{P(K)}{K} J_1(KR),
\end{equation}
using $dJ_0(x)/dx = J_1(x)$.

%% file: derivation_details/derivation_2.tex
Here we present the derivation of Equation \ref{eq:eQ}.

\noindent Using $L_2(\mu) = (3/2)\mu^2 - (1/2)$,
\begin{equation}
q_z^2 - \frac{q^2}{3} = q^2 \left(\mu_q^2 - \frac{1}{3}\right) = \frac{2q^2}{3}L_2(\mu_q)
\end{equation}
for a 3-d vector $\vec q$, 
and $L_\ell = \sqrt{4\pi/(2\ell+1)} Y_{\ell 0}$.
We note that 
\begin{equation}
\frac{3}{4}T_{zz} = 
\frac{1}{2}\int \frac{d^3k}{(2\pi)^3} L_2(\mu_k)\tilde\rho(\vec k) e^{i\vec k\cdot\vec r}.
\end{equation}
Finally, we have the expansion of a plane wave into spherical harmonics and
spherical Bessel functions:
\begin{equation}
e^{i\vec q\cdot\vec r} = 4\pi \sum_{\ell m} i^\ell j_\ell(qr) Y^*_{\ell m}(\hat q) Y_{\ell m}(\hat r).
\end{equation}

We then compute $\left<\epsilon_{zz} Q(r)\right>$ as
\begin{equation}
\left<\epsilon_{zz} Q(r)\right> = 
{5\tau}
\int\frac{d^3q}{(2\pi)^3} \frac{1}{2} L_2(\hat q) 
\int\frac{d^2\hat r}{4\pi} L_2(\hat r) 
\int\frac{d^3 k}{(2\pi)^3} e^{i\vec q\cdot\vec r} 
\left<\tilde\rho^*(\vec q)\tilde\rho(\vec k)\right>
\end{equation}
Converting to power, doing the $\vec k$ integral, and expanding the plane wave yields
\begin{equation}
\left<\epsilon_{zz} Q(r)\right> = 
\frac{5\tau}{2}
\int\frac{q^2 dq}{2\pi^2} P(q) \int \frac{d^2\hat q}{4\pi} L_2(\hat q) 
\int\frac{d^2\hat r}{4\pi} L_2(\hat r) 
4\pi \sum_{\ell m} i^\ell j_\ell(qr) Y^*_{\ell m}(\hat q) Y_{\ell m}(\hat r).
\end{equation}
We then can do the two angular integrals, yielding the simpler form:
\begin{equation}
\left<\epsilon_{zz} Q(r)\right> = 
-\frac{\tau}{2}
\int\frac{q^2 dq}{2\pi^2} P(q) j_2(qr).
\end{equation}

%% file: main.bbl
\begin{thebibliography}{}
\makeatletter
\relax
\def\mn@urlcharsother{\let\do\@makeother \do\$\do\&\do\#\do\^\do\_\do\%\do\~}
\def\mn@doi{\begingroup\mn@urlcharsother \@ifnextchar [ {\mn@doi@}
  {\mn@doi@[]}}
\def\mn@doi@[#1]#2{\def\@tempa{#1}\ifx\@tempa\@empty \href
  {http://dx.doi.org/#2} {doi:#2}\else \href {http://dx.doi.org/#2} {#1}\fi
  \endgroup}
\def\mn@eprint#1#2{\mn@eprint@#1:#2::\@nil}
\def\mn@eprint@arXiv#1{\href {http://arxiv.org/abs/#1} {{\tt arXiv:#1}}}
\def\mn@eprint@dblp#1{\href {http://dblp.uni-trier.de/rec/bibtex/#1.xml}
  {dblp:#1}}
\def\mn@eprint@#1:#2:#3:#4\@nil{\def\@tempa {#1}\def\@tempb {#2}\def\@tempc
  {#3}\ifx \@tempc \@empty \let \@tempc \@tempb \let \@tempb \@tempa \fi \ifx
  \@tempb \@empty \def\@tempb {arXiv}\fi \@ifundefined
  {mn@eprint@\@tempb}{\@tempb:\@tempc}{\expandafter \expandafter \csname
  mn@eprint@\@tempb\endcsname \expandafter{\@tempc}}}

\bibitem[\protect\citeauthoryear{Abareshi et~al.,}{Abareshi
  et~al.}{2022}]{abareshi_overview_2022}
Abareshi B.,  et~al., 2022, Technical report, Overview of the {Instrumentation}
  for the {Dark} {Energy} {Spectroscopic} {Instrument}, \url
  {https://ui.adsabs.harvard.edu/abs/2022arXiv220510939A}.
Lawrence Berkeley National Laboratory, \url
  {https://ui.adsabs.harvard.edu/abs/2022arXiv220510939A}

\bibitem[\protect\citeauthoryear{Anderson et~al.,}{Anderson
  et~al.}{2014}]{anderson_clustering_2014}
Anderson L.,  et~al., 2014, \mn@doi [Monthly Notices of the Royal Astronomical
  Society] {10.1093/mnras/stu523}, 441, 24

\bibitem[\protect\citeauthoryear{Binney}{Binney}{1985}]{binney_testing_1985}
Binney J.,  1985, \mn@doi [Monthly Notices of the Royal Astronomical Society]
  {10.1093/mnras/212.4.767}, 212, 767

\bibitem[\protect\citeauthoryear{Catelan \& Porciani}{Catelan \&
  Porciani}{2001}]{catelan_correlations_2001}
Catelan P.,  Porciani C.,  2001, \mn@doi [Monthly Notices of the Royal
  Astronomical Society] {10.1046/j.1365-8711.2001.04250.x}, 323, 713

\bibitem[\protect\citeauthoryear{Collaboration et~al.,}{Collaboration
  et~al.}{2020}]{collaboration_planck_2020}
Collaboration P.,  et~al., 2020, \mn@doi [Astronomy and Astrophysics]
  {10.1051/0004-6361/201833910}, 641, A6

\bibitem[\protect\citeauthoryear{{DESI Collaboration} et~al.,}{{DESI
  Collaboration} et~al.}{2016}]{desi_collaboration_desi_2016}
{DESI Collaboration} et~al., 2016, Technical report, The {DESI} {Experiment}
  {Part} {II}: {Instrument} {Design}, \url
  {https://ui.adsabs.harvard.edu/abs/2016arXiv161100037D}.
Lawrence Berkeley National Laboratory, \url
  {https://ui.adsabs.harvard.edu/abs/2016arXiv161100037D}

\bibitem[\protect\citeauthoryear{Dey et~al.,}{Dey
  et~al.}{2019}]{dey_overview_2019}
Dey A.,  et~al., 2019, \mn@doi [The Astronomical Journal]
  {10.3847/1538-3881/ab089d}, 157, 168

\bibitem[\protect\citeauthoryear{Gatti et~al.,}{Gatti
  et~al.}{2021}]{gatti_dark_2021}
Gatti M.,  et~al., 2021, \mn@doi [Monthly Notices of the Royal Astronomical
  Society] {10.1093/mnras/stab918}, 504, 4312

\bibitem[\protect\citeauthoryear{Gendzwill \& Stauffer}{Gendzwill \&
  Stauffer}{1981}]{gendzwill_analysis_1981}
Gendzwill D.~J.,  Stauffer M.~R.,  1981, \mn@doi [Journal of the International
  Association for Mathematical Geology] {10.1007/BF01031390}, 13, 135

\bibitem[\protect\citeauthoryear{Hadzhiyska, Eisenstein, Bose, Garrison  \&
  Maksimova}{Hadzhiyska et~al.}{2021}]{hadzhiyska_compaso_2021}
Hadzhiyska B.,  Eisenstein D.,  Bose S.,  Garrison L.~H.,   Maksimova N.,
  2021, \mn@doi [Monthly Notices of the Royal Astronomical Society]
  {10.1093/mnras/stab2980}, 509, 501

\bibitem[\protect\citeauthoryear{Hernquist}{Hernquist}{1990}]{hernquist_analytical_1990}
Hernquist L.,  1990, \mn@doi [The Astrophysical Journal] {10.1086/168845}, 356,
  359

\bibitem[\protect\citeauthoryear{Hirata}{Hirata}{2009}]{hirata_tidal_2009}
Hirata C.~M.,  2009, \mn@doi [Monthly Notices of the Royal Astronomical
  Society] {10.1111/j.1365-2966.2009.15353.x}, 399, 1074

\bibitem[\protect\citeauthoryear{Hirata \& Seljak}{Hirata \&
  Seljak}{2004}]{hirata_intrinsic_2004}
Hirata C.~M.,  Seljak U.,  2004, \mn@doi [Physical Review D]
  {10.1103/PhysRevD.70.063526}, 70, 063526

\bibitem[\protect\citeauthoryear{Hirata, Mandelbaum, Ishak, Seljak, Nichol,
  Pimbblet, Ross  \& Wake}{Hirata et~al.}{2007}]{hirata_intrinsic_2007}
Hirata C.~M.,  Mandelbaum R.,  Ishak M.,  Seljak U.,  Nichol R.,  Pimbblet
  K.~A.,  Ross N.~P.,   Wake D.,  2007, \mn@doi [Monthly Notices of the Royal
  Astronomical Society] {10.1111/j.1365-2966.2007.12312.x}, 381, 1197

\bibitem[\protect\citeauthoryear{Hogg \& Lang}{Hogg \&
  Lang}{2013}]{hogg_replacing_2013}
Hogg D.~W.,  Lang D.,  2013, \mn@doi [Publications of the Astronomical Society
  of the Pacific] {10.1086/671228}, 125, 719

\bibitem[\protect\citeauthoryear{Jackson}{Jackson}{1972}]{jackson_critique_1972}
Jackson J.~C.,  1972, \mn@doi [Monthly Notices of the Royal Astronomical
  Society] {10.1093/mnras/156.1.1P}, 156, 1P

\bibitem[\protect\citeauthoryear{Kaiser}{Kaiser}{1987}]{kaiser_clustering_1987}
Kaiser N.,  1987, \mn@doi [Monthly Notices of the Royal Astronomical Society]
  {10.1093/mnras/227.1.1}, 227, 1

\bibitem[\protect\citeauthoryear{Kitanidis et~al.,}{Kitanidis
  et~al.}{2020}]{kitanidis_imaging_2020}
Kitanidis E.,  et~al., 2020, \mn@doi [Monthly Notices of the Royal Astronomical
  Society] {10.1093/mnras/staa1621}, 496, 2262

\bibitem[\protect\citeauthoryear{Kong et~al.,}{Kong
  et~al.}{2020}]{kong_removing_2020}
Kong H.,  et~al., 2020, \mn@doi [Monthly Notices of the Royal Astronomical
  Society] {10.1093/mnras/staa2742}, 499, 3943

\bibitem[\protect\citeauthoryear{Landy \& Szalay}{Landy \&
  Szalay}{1993}]{landy_bias_1993}
Landy S.~D.,  Szalay A.~S.,  1993, \mn@doi [The Astrophysical Journal]
  {10.1086/172900}, 412, 64

\bibitem[\protect\citeauthoryear{Lang, Hogg  \& Mykytyn}{Lang
  et~al.}{2016}]{lang_tractor_2016}
Lang D.,  Hogg D.~W.,   Mykytyn D.,  2016, Astrophysics Source Code Library, p.
  ascl:1604.008

\bibitem[\protect\citeauthoryear{Maksimova, Garrison, Eisenstein, Hadzhiyska,
  Bose  \& Satterthwaite}{Maksimova et~al.}{2021}]{maksimova_abacussummit_2021}
Maksimova N.~A.,  Garrison L.~H.,  Eisenstein D.~J.,  Hadzhiyska B.,  Bose S.,
   Satterthwaite T.~P.,  2021, \mn@doi [Monthly Notices of the Royal
  Astronomical Society] {10.1093/mnras/stab2484}, 508, 4017

\bibitem[\protect\citeauthoryear{Martens, Hirata, Ross  \& Fang}{Martens
  et~al.}{2018}]{martens_radial_2018}
Martens D.,  Hirata C.~M.,  Ross A.~J.,   Fang X.,  2018, \mn@doi [Monthly
  Notices of the Royal Astronomical Society] {10.1093/mnras/sty1100}, 478, 711

\bibitem[\protect\citeauthoryear{Obuljen, Percival  \& Dalal}{Obuljen
  et~al.}{2020}]{obuljen_detection_2020-1}
Obuljen A.,  Percival W.~J.,   Dalal N.,  2020, \mn@doi [Journal of Cosmology
  and Astroparticle Physics] {10.1088/1475-7516/2020/10/058}, 2020, 058

\bibitem[\protect\citeauthoryear{Padilla \& Strauss}{Padilla \&
  Strauss}{2008}]{padilla_shapes_2008}
Padilla N.~D.,  Strauss M.~A.,  2008, \mn@doi [Monthly Notices of the Royal
  Astronomical Society] {10.1111/j.1365-2966.2008.13480.x}, 388, 1321

\bibitem[\protect\citeauthoryear{Singh, Mandelbaum  \& More}{Singh
  et~al.}{2015}]{singh_intrinsic_2015}
Singh S.,  Mandelbaum R.,   More S.,  2015, \mn@doi [Monthly Notices of the
  Royal Astronomical Society] {10.1093/mnras/stv778}, 450, 2195

\bibitem[\protect\citeauthoryear{Singh, Yu  \& Seljak}{Singh
  et~al.}{2021}]{singh_fundamental_2021}
Singh S.,  Yu B.,   Seljak U.,  2021, \mn@doi [Monthly Notices of the Royal
  Astronomical Society] {10.1093/mnras/staa3263}, 501, 4167

\bibitem[\protect\citeauthoryear{Tenneti, Mandelbaum, Di~Matteo, Feng  \&
  Khandai}{Tenneti et~al.}{2014}]{tenneti_galaxy_2014}
Tenneti A.,  Mandelbaum R.,  Di~Matteo T.,  Feng Y.,   Khandai N.,  2014,
  \mn@doi [Monthly Notices of the Royal Astronomical Society]
  {10.1093/mnras/stu586}, 441, 470

\bibitem[\protect\citeauthoryear{Troxel \& Ishak}{Troxel \&
  Ishak}{2015}]{troxel_intrinsic_2015}
Troxel M.~A.,  Ishak M.,  2015, \mn@doi [Physics Reports]
  {10.1016/j.physrep.2014.11.001}, 558, 1

\bibitem[\protect\citeauthoryear{Zhou et~al.,}{Zhou
  et~al.}{2021}]{zhou_clustering_2021}
Zhou R.,  et~al., 2021, \mn@doi [Monthly Notices of the Royal Astronomical
  Society] {10.1093/mnras/staa3764}, 501, 3309

\bibitem[\protect\citeauthoryear{Zhou et~al.,}{Zhou
  et~al.}{2022}]{zhou_target_2022}
Zhou R.,  et~al., 2022, Technical report, Target {Selection} and {Validation}
  of {DESI} {Luminous} {Red} {Galaxies}, \url
  {https://ui.adsabs.harvard.edu/abs/2022arXiv220808515Z}.
\url {https://ui.adsabs.harvard.edu/abs/2022arXiv220808515Z}

\makeatother
\end{thebibliography}
